# Automated Verification of Tree-Manipulating Programs Using Constrained Horn Clauses

MARCO FAELLA, University of Naples Federico II, Italy

GENNARO PARLATO, University of Molise, Italy

Verifying programs that manipulate tree data structures often requires complex, ad-hoc proofs that are hard to generalize and automate. This paper introduces an automatic technique for analyzing such programs. Our approach combines automata and logics to tackle the challenges posed by diverse tree data structures uniformly. At the core of our methodology is the *knitted-tree encoding*, which maps a program execution into a tree data structure encapsulating input, output, and intermediate configurations, within a single structure. By leveraging the compositional properties of knitted-trees, we characterize them using constrained Horn clauses (CHCs). This encoding reduces verification to solving CHC satisfiability, benefiting from ongoing advancements in CHC solvers. While we focus on the memory safety problem for illustration, our technique applies broadly to various verification tasks involving tree data structures.

## 1 Introduction

Automated analysis of programs that manipulate dynamic memory (heaps) remains a challenging problem in computer science, despite extensive research efforts. A major obstacle is the unpredictable nature of memory allocation and deallocation operations, which can make it hard to predict the program's state and the relationships between data elements over time. This unpredictability can lead to a state-space explosion, rendering exhaustive analysis computationally infeasible. Recursive procedures and unbounded loops introduce additional complexity, making reasoning about termination intricate. Additionally, the use of expressive logics for specifying invariants, preconditions, and postconditions adds another layer of complexity to the analysis task.

This paper introduces a novel technique for automated analysis of heap-manipulating programs, particularly those involving tree data structures. Our approach combines automata and logics to address these challenges. Ultimately, we reduce the verification to the satisfiability of constrained Horn clauses (CHCs). This reduction allows us to capitalize on advancements in CHC solvers, as demonstrated by the CHC-COMP competition [4, 17]. We illustrate our approach through its application to the *memory safety problem*. A program is deemed memory safe if no execution starting from a tree data structure that satisfies a given precondition (e.g., starting with a red-black tree) leads to crashes (such as *null-pointer dereferences*, *use after free*, *use of uninitialized memory*, and *illegally freeing memory*, as described in [32]) or results in non-termination.

At the core of our methodology lies the *knitted-tree encoding*, which maps an execution $\pi$ of a tree-manipulating program, starting from an input data tree $T$, into a tree data structure called the

Authors' Contact Information: Marco Faella, University of Naples Federico II, Italy, m.faella@unina.it; Gennaro Parlato, University of Molise, Italy, gennaro.parlato@unimol.it.

**knitted-tree** of $\pi$. A knitted-tree uniquely encapsulates the input, output, and all intermediate configurations of an execution within in a single structure. The underlying tree structure of the knitted-tree, or *backbone*, is derived from $T$ by adding a fixed number of initially inactive nodes to accommodate dynamic node allocation. Node labels consist in a sequence of records, called *frames*. In turn, frames are connected in a linear tree-wide sequence, called the *lace*, with consecutive frames belonging to the same node or to adjacent nodes of the tree, mimicking the knitting of a tree-shaped object. Each frame records changes to the node, including variables pointing to it, and the current program state, serving as a *log of updates*. This approach retains the original structure of the backbone while also representing the final heap, which may have a different graph structure. However, the parameters defining a knitted-tree – namely, the number of spare nodes added to the backbone and the number of frames in a node label – may not suffice to encompass all possible program executions, potentially leading to missed executions in our analysis.

Knitted-trees enjoy **compositional properties** that are essential for our verification approach using CHCs. Knitted-trees allow subtree replacement: we can substitute a subtree rooted at node $v$ in knitted-tree $T$ with a subtree from another knitted-tree rooted at $v'$, provided that the labels of $v$ and $v'$ satisfy an entirely local consistency property. Moreover, such consistency can be expressed in a quantifier-free first-order data theory. Using this replacement rule, we construct a CHC system whose minimal model precisely captures the set of valid node labels for the knitted-trees of a given program. This enables us to efficiently check for error configurations reachable along an execution. If our analysis does *not* find an error, we need an additional check to ensure the program's correctness. With a small change, the same CHC system can be used to verify if there are executions that cannot be represented using the current parameter values. In that case, we increment the parameters to capture more executions. As expected from the undecidability of the verification task, this process can potentially run indefinitely.

We can analyze a program under a given **precondition**, provided that the precondition can be expressed as a *Symbolic Data-Tree Automata* (SDTAS) [23]. SDTAS work bottom-up, starting at data tree leaves, using a tuple of state variables to represent the automaton's state and a separate set of variables for the alphabet. Transitions are defined as constraints in a joint multi-sorted first-order theory over the state and alphabet variables. SDTAS can characterize many common tree data structures, like binary-search trees (BST), red-black trees (RBT), AVL trees, and heaps [23]. Higher-level logic formalisms, like *Monadic Second-Order logic with Data* (MSO-D) [23], can naturally specify data tree properties and be translated into SDTAS. We chose SDTAS for their simplicity, which allows us to easily impose constraints on input tree data structures by constraining the backbone of the knitted-trees that naturally encode the input data trees. We use a construction similar to the cross-product construction for finite state automata. This construction can be naturally encoded by extending the system of CHCs characterizing the set of node labels of knitted-trees.

Our method also offers a unique approach to proving program termination. In knitted-trees, we attach at least one frame to a node for every instruction executed by the program. By limiting the number of frames per node, we can conclude that a program terminates for all inputs meeting the precondition if no knitted-tree exceeds the allowed number of frames at any node. While this condition might seem ad hoc, it applies to many procedures operating on well-known data structures, making our approach valuable for proving termination in many relevant cases.

Finally, we have implemented our approach in a prototype tool and used it to check the memory safety of some procedures involving well-known tree data structures. This demonstrates its practical feasibility in both producing proofs and discovering bugs. Additionally, we discuss how the proposed approach can be extended and generalized to verify more complex properties.

*Organization of the paper.* §2 covers basic notation and definitions including data trees and a generalization of SDTAs in §2.1, and CHCs in §2.2. §3 introduces a simple programming language for heap-manipulating programs and defines the memory safety problem. §4 defines the knitted-tree encoding, and its compositionality properties are detailed in §5. §6 describes our reduction to the CHC satisfiability problem, with program preconditions addressed in §7. § 8 presents a sound procedure for solving the memory safety problem. §9 demonstrates our approach's practicality, including our prototype implementation, optimizations, and experimental results. In §10, we discuss how to address other program properties using the framework presented in this paper. We discuss related work in §11 and conclude in §12, exploring potential future work and applications.

## 2 Preliminaries

This section outlines the basic notation and definitions used in the paper, including data trees and a generalization of SDTAs in §2.1, and CHCs in §2.2.

We denote the set of natural numbers by $\mathbb{N}$. For $i, j \in \mathbb{N}$ with $i \leq j$, we denote by $[i, j]$ the set of integers $k$ such that $i \leq k \leq j$. Also, we denote by $[j]$ the set $[1, j]$. We use the standard syntax and semantics for first-order logic (FOL) with equality [48]. Our framework includes formulas from a many-sorted quantifier-free first-order theory $\mathcal{D}$, encompassing various data types found in programs such as arithmetic, reals, and arrays. Hereafter, we refer to $\mathcal{D}$ as the *data theory*.

### 2.1 Data-Trees and Symbolic Data-Tree Automata

This section introduces data trees and symbolic data tree automata (SDTAs), which represent and characterize well-known tree data structures [23].

**Trees.** A $k$-ary tree $T$ is a finite, prefix-closed subset of $[k]^*$, where $k \in \mathbb{N}$. Each element in $T$ is a *node*, with the empty string $\varepsilon$ as the *root*. The tree has an implicit *edge relation*: for any $d \in [k]$, if both $v$ and $v.d$ are nodes in $T$, then $(v, v.d)$ is an *edge*, making $v.d$ the $d^{\text{th}}$ child of $v$, and $v$ the *parent* of $v.d$. Nodes without children are *leaves*, while those with children are *internal nodes*.

**Data signatures.** A *data signature* $\mathcal{S}$ is a collection of pairs $\{id_i : type_i\}_{i \in [n]}$, where each $id_i$ is a field name and $type_i$ is its data type (e.g., integers, floating point numbers, real numbers, Booleans $\mathbb{B}$, and fixed-length bit vectors). An *evaluation* $v$ of $\mathcal{S}$ maps each field name $id$ to its type-specific value, denoted $v.id$. The set of all evaluations, or the *language* of $\mathcal{S}$, is $L(\mathcal{S})$.

**Data-Trees.** A *data-tree* with data signature $\mathcal{S}$, or $\mathcal{S}$-*tree*, is a pair $(T, \lambda)$ where $T$ is a tree and $\lambda$ is a labeling function $\lambda : T \to L(\mathcal{S})$ that assigns an evaluation of $\mathcal{S}$ to each node $t \in T$. To simplify notation, the value of a field $id$ at node $t$ can be written as $t.id$ when $\lambda$ is clear from the context.

**Symbolic Data-Tree Automata.** *Symbolic data-tree automata* (SDTAs) extend traditional bottom-up finite tree automata for data trees. They use symbolic alphabets and states, defined by data signatures, with a transition function as a predicate from a many-sorted, quantifier-free first-order logic theory. Originally defined for binary trees [23], we extend SDTAs to $k$-ary trees.

*Definition 2.1.* A $k$-**ary symbolic data-tree automaton** $\mathcal{A}$ is a tuple $(\mathcal{S}^\Sigma, \mathcal{S}^Q, \psi, \psi^F)$ where:

- $\mathcal{S}^\Sigma$ is the *alphabet data signature* defining the tree alphabet $\Sigma = L(\mathcal{S}^\Sigma)$;
- $\mathcal{S}^Q$ is the *state data signature* defining the set of states $Q = L(\mathcal{S}^Q)$; $Q$ is extended to include **nil** for missing child nodes, forming $\widehat{Q} = Q \cup \{\textbf{nil}\}$.
- $\psi(q_1, q_2, \ldots, q_k, key, q)$ is the *transition constraint*, where $q_1, q_2, \ldots, q_k$ range over $\widehat{Q}$, $q$ ranges over $Q$, and $key$ is of type $\mathcal{S}^\Sigma$;
- $\psi^F$ is a unary constraint defining the set of *final states* $F \subseteq \widehat{Q}$, i.e., the set consisting of all elements $q \in Q$ such that $\psi^F(q)$ evaluates to *true*.

For a $k$-ary $\mathcal{S}^\Sigma$-tree $\mathcal{T} = (T, \lambda)$, a *run* of $\mathcal{A}$ on $\mathcal{T}$ is a function $\rho : T \to Q$ such that for all nodes $t \in T$, $\psi(q_1, q_2, \ldots, q_k, \lambda(t), \rho(t))$ holds, where $q_j$ is $\rho(t.j)$ if $t.j \in T$, otherwise **nil**. $(T, \lambda)$ is *accepted* by $\mathcal{A}$ if $T = \emptyset$ and $\psi^F(\textbf{nil})$ is *true*, or if there exists a run $\rho : T \to Q$ such that $\psi^F\big(\rho(\varepsilon)\big)$ is *true*. The *language* of $\mathcal{A}$, denoted $L(\mathcal{A})$, is the set of all accepted $\mathcal{S}^\Sigma$-trees.                          □

## 2.2 Constrained Horn Clauses

This section introduces the notation and definitions of constrained Horn clauses, or CHCs.

*Definition 2.2.* For a set $R$ of uninterpreted fixed-arity relation symbols representing the system's unknowns, a **constrained Horn clause** is a formula of the form $H \leftarrow C \wedge B_1 \wedge \cdots \wedge B_n$ where:

- $C$ is a *constraint* represented by a quantifier-free formula of the data theory $\mathcal{D}$ that does not contain any relation symbol from $R$;
- for every $i \in [n]$, $B_i$ is an application $r(v_1, \ldots, v_k)$ of a relation symbol $r \in R$ to first-order variables $v_1, \ldots, v_k$;
- $H$ is the clause *head* and is either *false*, or it is an application $r(v_1, \ldots, v_k)$ of a relation symbol $r \in R$ to the first-order variables $v_1, \ldots, v_k$.

A CHC is a *fact* if its body has only the $C$ component, and it is a *query* if its head is *false*. A finite set $C$ of CHCs is called a *system*, corresponding to the first-order formula obtained by conjoining all its CHCs, with all free variables universally quantified. We assume that the semantics of constraints is given a priori as a structure.                          □

A CHC system $\mathcal{S}$ with relation symbols $R$ is *satisfiable* if there exists an interpretation for each relation symbol in $R$ that makes all clauses in $\mathcal{S}$ valid. The CHC *satisfiability problem* is the computational task of determining whether a given system $\mathcal{S}$ of CHCs is satisfiable.

Every system of CHCs $\mathcal{S}$ has a unique minimal model w.r.t. the subset ordering,[1] computable as the fixed-point of an operator derived from the clauses of $\mathcal{S}$ [37, 66]. This fixed-point semantics justifies the correctness of the reductions in this paper.

## 3 Heap-manipulating Programs and the Memory Safety Problem

This section introduces a simple programming language with specialized heap manipulation functionality and defines the memory safety problem for such programs.

The *heap* is a fundamental data structure for dynamic memory allocation, allowing memory blocks (or *nodes*) to be allocated and deallocated during program execution. These nodes typically have a *data field* and pointer fields linking them. Formally, a *specific heap state*, is defined as follows.

*Definition 3.1.* A **heap** is a tuple $\mathcal{H} = (\mathcal{S}, N, \textbf{data}, PF)$ where

- $\mathcal{S}$ is a data signature defining the type of data that can be stored in a node.
- $N$ is a finite set of nodes, including a unique element **nil** representing unallocated memory.
- $\textbf{data} : (N \cup \{\textbf{nil}\}) \to L(\mathcal{S})$ assigns a data signature evaluation of $\mathcal{S}$ to each node $u \in N$, except for **nil**. modeling the data field of each node.
- $PF$ is a finite sequence of distinct named pointer fields representing the pointer fields of each node: each pointer field is a function mapping nodes except **nil** to another heap node, i.e., of type $(N \setminus \{\textbf{nil}\}) \to N$.                          □

**Syntax.** Our programming language syntax is shown in Fig. 1a. A program begins with declarations of pointer and data variables, followed by a finite sequence of labeled instructions. Pointer variables can point to heap nodes, or **nil**, and data variables store values within their data type range. Instructions include assignments, control flow, and heap operations. Data assignments are of the

---

[1]See [66] for logic programs and Proposition 4.1 in [37] for constrained logic programs, equivalent to CHCs.

$$\text{Program} \stackrel{\text{def}}{=} decl\ pc\_stmt^+$$

$$decl \stackrel{\text{def}}{=} (\textbf{pointer}\ id(,id)^*)^*\ (type\ id(,id)^*)^*$$

$$pc\_stmt \stackrel{\text{def}}{=} pc : stmt ;$$

$$stmt \stackrel{\text{def}}{=} ctrl\_stmt \mid heap\_stmt$$

$$ctrl\_stmt \stackrel{\text{def}}{=} d := exp \mid d_{\text{bool}} := heap\_cond \mid \textbf{skip} \mid \textbf{exit}$$
$$\mid \textbf{if}\ cond\ \textbf{then}\ pc\_stmt^*\ \textbf{else}\ pc\_stmt^*\ \textbf{fi}$$
$$\mid \textbf{while}\ cond\ \textbf{do}\ pc\_stmt^+\ \textbf{od}$$

$$heap\_stmt \stackrel{\text{def}}{=} \textbf{new}\ p \mid \textbf{free}\ p$$
$$\mid p := \textbf{nil} \mid p := q \mid p := q {\rightarrow} pfield$$
$$\mid p {\rightarrow} pfield := \textbf{nil} \mid p {\rightarrow} pfield := q$$
$$\mid p {\rightarrow} \textbf{data} := exp \mid d := p {\rightarrow} \textbf{data}$$

$$exp \stackrel{\text{def}}{=} d \mid f(exp, \ldots, exp)$$

$$cond \stackrel{\text{def}}{=} r(exp, \ldots, exp) \mid (\neg)?\ heap\_cond$$

$$heap\_cond \stackrel{\text{def}}{=} p = q \mid p = \textbf{nil}$$
$$\mid p {\rightarrow} pfield = q \mid p {\rightarrow} pfield = \textbf{nil}$$

(a) Syntax of a basic programming language over a heap signature. *pfield* is the name of one of the pointer fields (i.e., one of the functions in *PF*). *p* and *q* are variables of type pointer; *d* is a variable of data type; *f* is a function symbol from the data theory and *r* is a relation symbol from the data theory; *pc* is the instruction counter; *type* is one of the sorts from the underlying data theory.

```
pointer root, p, p′
int key, value
 0 :  p′ := root ;
 1 :  while (p′ ≠ nil) do
 2 :      value := p′→data ;
 3 :      p := p′ ;
 4 :      if (key ≤ value) then
 5 :          p′ := p′→left ;
          else
 6 :          p′ := p′→right ;
          fi ;
      od ;
 7 :  new p′ ;
 8 :  p′→data := key ;
 9 :  if (root = nil) then
10 :      root := p′ ;
      else
11 :      if (key ≤ value) then
12 :          p→left := p′ ;
          else
13 :          p→right := p′ ;
          fi ;
      fi ;
14 :  exit ;
```

(b) Bst Insertion.

Fig. 1. Basic programming language for heap-manipulating programs.

form $d := exp$, where $d$ is a data variable set to the value of the data expression $exp$. Data expressions are constructed from data variables and combined using function symbols of the data theory $\mathcal{D}$. Control flow instructions include **skip**, **exit**, conditionals (**if-then-else**), and loops (**while**). Boolean conditions (*cond*) are either *data conditions* or *heap conditions*, not both simultaneously, though this can be managed by introducing Boolean variables for each condition. Heap conditions can be assigned to a Boolean variable with $d_{\text{bool}} := heap\_cond$, integrating them into Boolean theory. Heap operations include creating a new node with **new** $p$ (assigning $p$ to the new node and initializing its fields to undefined values or **nil**) and deallocating a node with **free** $p$ (setting all pointers to the node to **nil**). We also allow assignment and retrieval of values from node fields.

*Handling recursion:* Our language lacks direct support for function calls. We can handle non-recursive procedure calls by inlining its code at call points. We can also handle recursive functions, such as those used for manipulating tree data structures. These functions typically operate on a single node at a time, recursively calling themselves on the children of that node. To simulate this type of recursion, we can add a pointer field to each node that points to the node from which the current function call was made, allowing navigation back to the previous node.

A program $P$ is **valid** if it is well-formed, type-correct, has unique labels for each statement, and ends with an **exit** statement. Fig. 1b shows a program that inserts the value stored in *key* into a binary search tree (Bst). In $P$, $PC_P$, $PV_P$, and $DV_P$ are sets of program counters, pointer variables, and data variables, respectively. The function $succ : PC_P \rightharpoonup PC_P$ determines the successor of a program counter, if one exists. The **exit** statement has no successor, while control-flow statements have two successors based on a Boolean condition, denoted by $succ(pc, true)$ and $succ(pc, false)$.

**Semantics.** A program $P$ operates on a specialized heap called a *$P$-heap*, containing all the pointers and all the data and pointer fields in $P$. A *configuration* of $P$ is a tuple $(\mathcal{H}, v_p, v_d, pc)$ consisting of a $P$-heap, an evaluation of the pointer variables in $P$, an evaluation of the data variables in $P$, and a label indicating the next instruction to be executed. Since we focus on programs operating on tree data structures, a configuration $c$ is an *initial configuration* when satisfies the following conditions:

- $\mathcal{H}$ is *isomorphic* to a data tree $\mathcal{T}$ via a bijection $\rho$ that maps each node in $\mathcal{H}$ to a node in $\mathcal{T}$, such that for all nodes $x, y$ in $\mathcal{H}$ and $i \in |PF|$, $y = pf_i(x)$ iff $\rho(y) = \rho(x).i$, where $pf_i$ is the $i$-th pointer field in $PF$. We refer to $\mathcal{T}$ as the data tree of $c$, and may use $\mathcal{T}$ in place of $\mathcal{H}$.
- $v_p$ maps the first pointer variable declared in $P$ to the $\mathcal{H}$ node corresponding to the root of $\mathcal{T}$ and maps the other pointer variables to **nil**.
- $v_d$ assigns each variable a non-deterministic value.
- $pc$ is the label of the first statement in $P$.

A *transition* in $P$ from configuration $c$ to $c'$, denoted $c \rightarrow_P c'$, occurs when the instruction at $pc$ is executed, following standard instruction semantics unless noted otherwise. If $pc$ corresponds to an **exit** statement, $c$ is a *final configuration*, with no further transitions. If a **nil** pointer is dereferenced or used for node deallocation, the transition does not occur, making $c$ an *error configuration*. An **execution** $\pi$ of $P$ is a potentially infinite sequence of configurations $c_0 c_1 \ldots$ of $P$, where: *(i)* $c_0$ is initial, and *(ii)* $c_{i-1} \rightarrow_P c_i$ is valid for each configuration index $i \in \mathbb{N}$. A finite $\pi$ ending in a final or error configuration is a *terminating* or *buggy execution*. $\pi$ complies with a precondition given by an Sdta $\mathcal{A}_{pre}$ if a data tree $\mathcal{T}_0$ isomorphic to the $P$-heap in $c_0$, belongs to the language of $\mathcal{A}_{pre}$.

**Memory safety problem.** A program $P$ is *memory safe* with respect to a precondition $\mathcal{A}_{pre}$ if all its executions complying with $\mathcal{A}_{pre}$ are terminating, meaning no execution is or end in an error configuration. The *memory safety problem* asks whether $P$ is memory safe with respect to a given precondition $\mathcal{A}_{pre}$. It is simple to prove that the memory safety problem is undecidable.

## 4 Knitted-trees: a data tree representation of executions

Our procedure for solving the memory safety problem is based on the knitted-tree encoding, which we present in this section. The *knitted-tree* encoding represents each program execution as a data tree, capturing input, output, and intermediate configurations, facilitating the verification of preconditions, (postconditions,) and other specifications of interest within a unified structure.

Before describing the encoding, we introduce the necessary notation and review the assumptions and parameters involved. The encoding uses two parameters, $m, n \in \mathbb{N}$, which will be explained later. Assume $\pi$ is an execution of $P$ starting from an initial configuration $c_0$, where $\mathcal{T} = (T, \lambda)$ is a $k$-ary data tree of $c_0$ with signature $\mathcal{S} = \{key : \mathcal{D}\}$, and $\widehat{p}$ points to the root of $\mathcal{T}$ in $c_0$. The encoding maps $\pi$ to a data tree $\mathcal{K} = (K, \mu)$, called the *$(m, n)$-knitted-tree* of $\pi$. For technical reasons, we adopt a non-deterministic encoding, meaning that an execution $\pi$ can lead to a set of $(m, n)$-knitted-trees, denoted by $kt(\pi, m, n)$. The rationale behind this non-deterministic approach is detailed in §4.3. We now describe a generic knitted-tree $\mathcal{K} = (K, \mu)$ from the set $kt(\pi, m, n)$.

### 4.1 The backbone

The tree $K$, also called the *backbone* of $\mathcal{K}$, is the smallest tree satisfying the following conditions:

(1) The input tree $T$ is a subset of $K$ ($T \subset K$).
(2) All nodes from $T$ are internal nodes in $K$.
(3) The degree of all internal nodes in $K$ is $k + m$.
(4) There is at least one internal node.

Note that, in the special case where $T$ is empty, the backbone $K$ is a full tree of height 2, consisting of a root and its $k + m$ children. Fig. 2 illustrates a backbone structure.

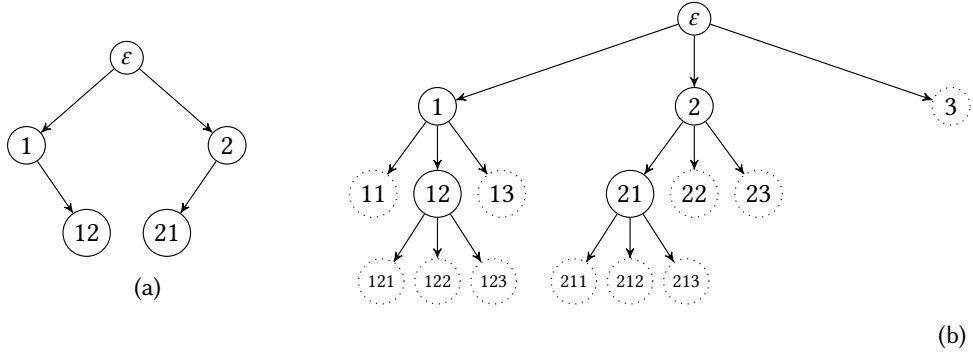

(a)

(b)

Fig. 2. **Example of backbone:** Part (b) illustrates a backbone derived from the input tree displayed in Part (a). Here, we assume $k = 2$ and $m = 1$.

Intuitively, each node in $K$ represents a distinct heap node. Initially, all nodes in $T$ are active, while the remaining nodes are inactive (dotted in Fig. 2). When an active node is freed using the **free** instruction, it transitions into an inactive state, as explained later in this section.

### 4.2 The node log

The backbone of a knitted-tree is determined solely by the input data tree and the parameters $m$ and $n$, regardless of the execution it represents. We now annotate the backbone to track modifications to the input data tree along $\pi$ by labeling the nodes. Each node is equipped with a sequence of *frames*, called the *log* of the node. Each frame typically documents a specific modification related to the node, such as altering the data it contains or changing one of its pointer fields, or assigning that node to a pointer variable. To establish connections between nodes and create a chronological record of updates, we organize the frames into a doubly linked list that enables us to navigate the execution in both directions. Additionally, if two frames correspond to consecutive operations located in different nodes, we add frames to the nodes along the path that connects two nodes in the backbone. Frames also contain information about the ongoing program state. The data signature $\mathcal{S}_{\mathcal{K}}$ of the knitted-tree $\mathcal{K}$ specifies the structure of each node's log in $\mathcal{K}$, as illustrated in Figure 3.

A *log* is a sequence of $n + 1$ frames, each identified by its index $i$. Frames within the same node are sorted temporally by index. Once a specific frame $f$ of a log $\sigma$ is identified and named, we omit its index $i$ when referring to its fields (e.g., $f.pc$ instead of $\sigma.pc^i$). The last frame within each label handles *label overflow*, which occurs when more than $n$ frames are needed in a node, requiring the program to use the $(n + 1)$th frame. The *prev* field holds a value from $Dir \times [n]$, not $Dir \times [n + 1]$, since a frame with index $n + 1$ has no successor. The *instr* field of a frame contains the *events*, also referred to as *symbolic instructions*, which are categorized as follows:

$$Instr = \big\{ \langle pfield := p \rangle, \langle pfield := \mathbf{nil} \rangle, \langle p := \mathbf{here} \rangle \mid pfield \in PF, p \in PV_P \big\}$$

$$\cup \ \big\{ \text{RWD}_i \mid i \in [n] \big\} \cup \big\{ \text{RWD}_{i,p} \mid i \in [n], p \in PV_P \big\} \ \cup \ \big\{ \text{NOP}, \text{ERR}, \text{OOM} \big\} \ .$$

$$\mathcal{S}_{\mathcal{K}} = \left\{\begin{array}{ll} avail^i : \mathbb{B}, & \text{Is this frame available?} \\[4pt] active^i : \mathbb{B}, & \text{Is this node allocated?} \\[4pt] key^i : \mathcal{D}, & \text{Current value of this node's data field} \\[4pt] pc^i : PC_P, & \text{Program counter} \\[4pt] \{d^i : \mathcal{D}_d\}_{d \in DV_P}, & \text{Current value of the data variables} \\[4pt] \{upd^i_p : \mathbb{B}\}_{p \in PV_P}, & \text{Has } p \text{ been updated since the frame } i-1? \\[4pt] \{isnil^i_p : \mathbb{B}\}_{p \in PV_P}, & \text{Is } p \text{ } \mathbf{nil}? \\[4pt] instr^i : Instr, & \text{Symbolic instruction} \\[4pt] active\_child^i : \mathbb{B}^{k+m}, & \text{Is each child allocated?} \\[4pt] next^i : Dir \times [2, n+1], & \text{Link to the next frame} \\[4pt] prev^i : Dir \times [2, n] & \text{Link to the previous frame} \end{array}\right\}_{i \in [n+1]},$$

where $Dir = \{-, \uparrow\} \cup [k+m]$ encodes the position of an adjacent frame relative to the frame node.

Fig. 3. Data signature $\mathcal{S}_{\mathcal{K}}$ of the knitted-tree $\mathcal{K}$.

The first group of events logs refers to changes in the node's pointer fields, maintaining a chronological record of the pointer variables referencing the node. The events $\textsc{rwd}_i$ and $\textsc{rwd}_{i,p}$ represent a *lace rewinding* operation, crucial for the sequencing node updates. Other symbols include: an empty instruction ($\textsc{nop}$), a null-pointer dereference ($\textsc{err}$), and an out-of-memory error ($\textsc{oom}$) resulting from excessive use of the statement **new**. Detailed explanations of these events will follow.

### 4.3 The labeling function

The labeling function $\mu$ of $\mathcal{K}$ is defined inductively on the length of $\pi$.

*4.3.1* **Base case.** $\pi$ consists of an initial configuration, say $(\mathcal{T}, v_p, v_d, pc)$. We encode the input tree $\mathcal{T} = (T, \lambda)$ into the backbone by setting the first frame of each node $t \in K$ as follows:

$$\mu(t).avail^1 = false, \qquad\qquad \mu(t).active^1 = \begin{cases} true & \text{if } t \in T \\ false & \text{otherwise} \end{cases}$$

$$\mu(t).key^1 = \begin{cases} \lambda(t).key & \text{if } t \in T \\ \text{unspecified} & \text{otherwise} \end{cases} \qquad \mu(t).active\_child^1_j = \begin{cases} true & \text{if } t.j \in T \\ false & \text{otherwise.} \end{cases}$$

The other fields of the first frame are unspecified (i.e., they assume all possible values in different knitted-trees). The initial configuration of $\pi$ is stored in the root's *second* frame: $\mu(\varepsilon).avail^2 = false$, $\mu(\varepsilon).prev^2 = (-, 2)$ (a self-loop), $\mu(\varepsilon).pc^2$ is set to the first statement's label in $P$, and $\mu(\varepsilon).isnil^2_q = true$ for all pointer variables $q$ different from $\widehat{p}$. If $T$ is not empty, $\mu(\varepsilon).instr^2 = \langle \widehat{p} := \mathbf{here} \rangle$ and $\mu(\varepsilon).isnil^2_{\widehat{p}} = false$; otherwise, $\mu(\varepsilon).instr^2 = \textsc{nop}$ and $\mu(\varepsilon).isnil^2_{\widehat{p}} = true$. All remaining frames in all nodes are labeled as available. The fields *active*, *key*, and *active\_child* of the second frame of the root are copied from the root's first frame.

*4.3.2* **Inductive case.** We start with an overview of the encoding method, its properties, and the required notation. Let $\pi = \overline{\pi}c$, where $c$ is a configuration and $\overline{\pi}$ is a non-empty execution. Assume that $\overline{\mathcal{K}} = (K, \overline{\mu})$ is a knitted-tree in $kt(\overline{\pi}, m, n)$. We define the labeling $\mu$ for $\pi$ by extending the lace of $\overline{\mathcal{K}}$ based on the last instruction executed in $\pi$. To aid understanding, we list some invariants for all knitted-trees, providing an informal explanation for brevity.

*The lace.* Besides individual node logs, we maintain a chronological order of all frames across all nodes to track the execution sequence of $\pi$. All the unavailable frames in the knitted-tree with index greater than 1 form a doubly linked list called the *lace* using the *next* and *prev* fields of the frames. The first frame in the lace is the root's second frame. We identify a frame in a knitted-tree by a pair $(u, i)$, where $u$ is a node and $i \in [n+1]$ is the frame's index. We say that frame $(v, j)$ is the *lace successor* of frame $(u, i)$ and write $(u, i) \rightarrow_{next} (v, j)$ (and $(u, i)$ is the *lace predecessor* of $(v, j)$, written $(v, j) \rightarrow_{prev} (u, i)$) if $i, j > 1$ and one of the following cases holds:

- $u = v$, $j = i + 1$, $\mu(u).next^i = (-, j)$, and $\mu(v).prev^j = (-, i)$;
- $v$ is the $l^{\text{th}}$ child of $u$, $\mu(u).next^i = (l, j)$, and $\mu(v).prev^j = (\uparrow, i)$;
- $u$ is the $l^{\text{th}}$ child of $v$, $\mu(u).next^i = (\uparrow, j)$, and $\mu(v).prev^j = (l, i)$.

Available frames' unspecified fields contribute to $kt(\pi, m, n)$'s non-determinism.

*Properties of the frame fields.* In our inductive definition, we extend the knitted-tree for $\pi$ from one defined for $\overline{\pi}$ by appending frames to the nodes' logs based on $\pi$'s last transition. This helps us assign meanings to fields like $upd_p$, $isnil_p$, $d$, $key$, $pc$, and $active\_child$ for the unavailable frames. The $upd_p$ flag tracks changes to the pointer $p$ outside the current node: it is set to *true* in frame $(u, i)$ (for $i > 1$) if $p$ was assigned a non-**nil** value in the part of the lace between frames $(u, i-1)$ and $(u, i)$, excluding these frames. Thus, if frames $(u, i-1)$ and $(u, i)$ are adjacent in the lace (i.e., $(u, i-1) \rightarrow_{next} (u, i)$, aka a *local step*), all $upd_p$ flags in $(u, i)$ are *false*. The $isnil_p$ flag is *true* in a frame if $p$ equals **nil** at that point in the execution. The $active\_child$ flags help track the allocation of the child nodes of the backbone. Other fields maintain their original meanings in the configuration.

*Pushing a frame.* Appending a frame to a log involves: (1) finding the smallest index $i$ where $avail^i$ is *true*, and (2) adding the new frame at position $i$. Therefore, a log operates akin to a stack, with the bottom frame at index 1 and the *top frame* being the highest-index frame where $avail$ is *false*.

*Default values for a frame.* Any frame pushed onto a log assumes default values unless specified otherwise. When pushing a frame $f$ on a node $u$, default values come from the preceding frame in the lace, $f^{\text{prev}}$, and the frame below $f$ in $u$'s log, $f^{\text{below}}$. Note that $f^{\text{prev}}$ and $f^{\text{below}}$ can be the same. The default values for the fields of $f$ are as follows: for all $p \in PV_P$ and $d \in DV_P$,

- $avail = false$, $instr = \text{NOP}$, and $upd_p = false$;
- $active$ and $key$ are copied from $f^{\text{below}}$;
- $isnil_p$, $d$, and $pc$ are copied from $f^{\text{prev}}$;
- $active\_child_j$ is copied from $f^{\text{prev}}.active$ if $f^{\text{prev}}$ belongs to the $j^{\text{th}}$ child of $u$; otherwise, it is copied from $f^{\text{below}}.active\_child_j$;
- $prev$ points to $f^{\text{prev}}$;
- $next$ is unspecified and can take any value in different knitted-trees for the same execution.

Moreover, $f^{\text{prev}}.next$ is updated to point to $f$. This resolves the non-determinism left in the *next* field of the previous frame, in such a way that it points to the newly added frame.

Despite the non-determinism in the *next* field, identifying the last frame $f$ in a lace can be done by checking $f$ and its lace successor $f'$. Specifically, $f$ is the last frame of the lace if $f'$ is available or $f'$ precedes $f$ in the lace, which happens when field of $f'.prev$ does not point to $f$.

Henceforth, we assume that $\overline{f}$ is the last frame of the lace in $\overline{\mathcal{K}}$ located as top frame of node $\overline{t}$.

Instructions are classified as either *local* or *walking*. Local instructions only use information from $\overline{f}$, and push a single new frame $f$ onto $\overline{t}$. The fields of $f$ are set to default values, except for those specified below. The following cases correspond to the instruction pointed to by $\overline{f}.pc$:

**Encoding of $p \coloneqq$ nil:** $f.pc = succ(\overline{f}.pc)$ and $f.isnil_p = true$.

**Encoding of $d \coloneqq exp$:** $f.pc = succ(\overline{f}.pc)$ and $f.d$ is set to the value of $exp$, with variables in $DV_P$ evaluated using their values from $\overline{f}$.

**Encoding of skip:** $f.pc = succ(\overline{f}.pc)$.

Non-local or *walking* instructions may access different nodes in addition to $\overline{t}$. The main reasons to move to another node are to dereference a pointer or identify the node a pointer field refers to. To get this information, we *rewind the lace* by moving backward to find the most recent assignment to the relevant pointer variable. For example, to identify the node a pointer variable $p$ points to, we rewind the lace until we find a frame with the symbolic instruction $\langle p \coloneqq \text{here} \rangle$.

*Lace rewinding path.* To define the process of rewinding the lace of $\overline{\mathcal{K}}$, we introduce an auxiliary function called $find\_ptr(p, id)$. This function takes two parameters: a pointer variable $p \in PV_P$, and a frame identifier $id = (u, i)$. It returns a sequence of frame ids by tracing the lace backward from $id$, using optimizations to take shortcuts when possible. Instead of following the lace step-by-step, we move backward in the lace only if we are sure to find an update to $p$ before returning to the current node. We use the $upd_p$ flag in the current frame to check this: if *true*, we go backward, otherwise, we take a shortcut by moving down in the log. Specifically, we jump directly to the topmost frame with $upd_p = true$, considering it in the next step. If no such a frame exists, we jump to the frame in position 2 of the current log (the frame in position 1 is special and is not part of the lace). Formally, $find\_ptr(p, id)$ is the sequence of frame ids $id_1, \ldots, id_l$ where $id_j = (u_j, i_j)$ for all $j \in [l]$, such that: *(i)* $id_1 = id$; *(ii)* for all $j \in [l-1]$, $id_j \rightarrow_{prev} id_{j+1}$ if $\overline{\mu}(u_j).upd_p^{i_j} = true$, otherwise $u_{j+1} = u_j$ and $i_{j+1}$ is the maximum value between 2 and the greatest index less than $i_j$ such that $\overline{\mu}(u_j).upd_p^{i_{j+1}} = true$; *(iii)* the frame with id $id_l$ contains the symbolic assignment $\langle p \coloneqq \text{here} \rangle$; and *(iv)* all previous frames do not contain symbolic assignments to $p$.

*Example 4.1.* We illustrate the rewinding operation with an example. Consider the fragment of knitted-tree in Fig. 4, showing node $u$ and its children $v_1$ and $v_2$. Each node displays a small selection of the information contained within each frame. Gray numbers *above* the frames indicate label positions, while numbers *inside* the frames represent lace positions (*ordinals*). The lace starts at the frame with ordinal 1 of the root (node $u$), takes a local step to the frame with ordinal 2, then moves to the left child $v_1$, and so on. Suppose that the program counter in the 7th frame points to the statement $p \coloneqq q$, that dereferences pointer $q$. Using 7 as a shorthand for the frame $(v_2, 2)$, we apply $find\_ptr(q, 7)$ to find the sequence of frames to rewind to the most recent assignment to $q$. In terms of ordinals, such sequence is 7, 6, 4, 3. In terms of frame ids we have:

$$find\_ptr(q, (v_2, 2)) = (v_2, 2), (u, 6), (u, 4), (v_1, 2).$$

Notice that the sequence skips the frame with ordinal 5 because it contains $upd_q = false$. Frames with ordinals 8 and 9 are explained later, in Example 4.2. □

*Null pointer dereference.* If the instruction located at $\overline{f}.pc$ dereferences a pointer $p$, and $p$ is **nil**, it indicates an error. Thus, if the flag $isnil_p$ is *true* in $\overline{f}$, we push a new frame with the symbolic instruction ERR onto the current node $\overline{t}$ to indicate a runtime error. For the lace construction, we assume that dereferenced pointers are not **nil** when the last executed instruction of $\pi$ is executed.

**Encoding of $p \rightarrow pfield \coloneqq$ nil.** We rewind the lace to find the most recent assignment to $p$. Let $(\overline{t}, \overline{i})$ be the identifier of $\overline{f}$, and $id_1 = (u_1, i_1), \ldots, id_l = (u_l, i_l)$ be the sequence $find\_ptr(p, (\overline{t}, \overline{i}))$. In principle, we might perform the encoding by pushing a new frame for each element of the sequence $id_1, \ldots, id_l$, to keep a faithful record of the movements necessary to simulate the current statement. Instead, we apply the following optimizations, whose objective is to use fewer frames:

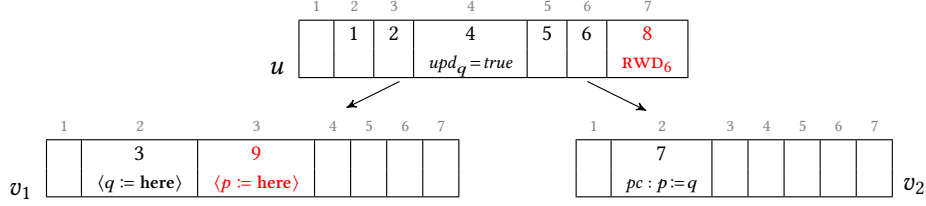

Fig. 4. A 3-node fragment of a knitted-tree with parameters $k = 2$, $n = 7$, $m = 0$. The extra frame of position $n + 1$ is not depicted. Gray numbers above the frames represent positions within the label, whereas numbers inside the frames represent their positions in the lace.

*(i)* we skip the first id $(u_1, i_1)$ because it is redundant; *(ii)* for each of the following ids $id_j$ in the sequence, we only push a new frame if $id_{j-1}$ belongs to a different node than $id_j$. Precisely, for all $j = 2, \ldots, l - 1$, if $u_j \neq u_{j-1}$ then we push a frame onto $u_j$ with $instr = \text{RWD}_{i_j}$. The program counter does not advance on these frames. Then, for the last id of the sequence, we push a frame $f$ onto $u_l$ with $f.pc = succ(\overline{f}.pc)$ and $f.instr = \langle \textit{pfield} := \textbf{nil} \rangle$. The returned node is $u_l$, i.e., $u_l = t$.

**Encoding of $p \rightarrow \textit{pfield} := q$.** The encoding process is the same as that described for $p \rightarrow \textit{pfield} := \textbf{nil}$, with one key difference: in the final frame, we set the symbolic instruction to $\langle \textit{pfield} := q \rangle$.

**Encoding of $p := q \rightarrow \textit{pfield}$.** First, the lace moves to the node $u$ pointed by $q$ by following the rewinding process described above for the statement $p \rightarrow \textit{pfield} := \textbf{nil}$. Then, we search in $u$'s log for the most recent assignment to the *pfield* field. This operation is not done by going backward in the lace, but rather by going backward *in the log* of $u$, since it must be recorded in $u$'s log per our encoding invariant. Specifically, we search for the largest index $i$ where the frame $(u, i)$ is in use and contains a symbolic instruction of the form $\langle \textit{pfield} := \alpha \rangle$, for some $\alpha$. If no such index exists, *pfield* is interpreted as having its default value, which is to point at one of the children of $u$ in the original input tree. We then distinguish cases:

**[$\alpha = \textbf{nil}$]** We push a frame with $isnil_p = \textit{true}$ on $u$.

**[$\alpha = r$, for some $r \in PV_P$]** If $isnil_r$ is *true* in the current frame, the lace pushes a frame with $isnil_p = \textit{true}$. Otherwise, the lace moves again to the node pointed by $r$, and there it pushes a frame with $\langle p := \textbf{here} \rangle$.

**[The log of $u$ does not contain an explicit assignment to $\textit{pfield}$]** If *pfield* is the $j^{\text{th}}$ field in *PF*, $u.j \in T$, and $u.j$ is active (as encoded in the flag $active\_child_j$), the lace moves to $u.j$ and pushes a frame with symbolic instruction $\langle p := \textbf{here} \rangle$ there. Otherwise, a frame with $isnil_p = \textit{true}$ is pushed on $u$.

The last pushed frame always updates the program counter, irrespective of the executed instruction.

**Encoding of $p := q$.** If $isnil_q$ evaluates to *true* in $\overline{f}$, the lace pushes a new frame onto $\overline{t}$ with $isnil_p$ set to *true*. Otherwise, the lace moves to the node pointed by $q$ and pushes a frame with $\langle p := \textbf{here} \rangle$ into it. In either case, the last frame also updates the program counter ($pc$).

*Example 4.2.* Consider again the knitted-tree in Fig. 4. Recall that the program counter in frame $(v_2, 2)$ points to the statement $p := q$. As explained in Example 4.1, $find\_ptr(q, (v_2, 2))$ returns the sequence of ids $(v_2, 2)$, $(u, 6)$, $(u, 4)$, $(v_1, 2)$. Due to the optimizations described above, the encoding of the current statement adds only two frames to the lace, with ordinals 8 and 9. Frame 8 does not advance the pc and carries the special symbolic instruction $\text{RWD}_6$, that keeps track of the position, in the current log, reached by the rewinding process. Frame 9, instead, advances the pc and encodes the actual assignment to $p$, in the same node where $q$ currently points.                    □

| Statement | Movement | Information stored |
|---|---|---|
| $p \rightarrow pfield \coloneqq \mathbf{nil}$ | Find $p$ or fail | $\langle pfield \coloneqq \mathbf{nil} \rangle$ |
| $p \rightarrow pfield \coloneqq q$ | Find $p$ or fail | $\langle pfield \coloneqq q \rangle$ or $\langle pfield \coloneqq \mathbf{nil} \rangle$ |
| $p \coloneqq q$ | Find $q$ | $\langle p \coloneqq \mathbf{here} \rangle$ or set $isnil_p$ |
| $p \coloneqq q \rightarrow pfield$ | Find $q$ or fail, then find last assignment to $pfield$ | $\langle p \coloneqq \mathbf{here} \rangle$ or set $isnil_p$ |
| $p \rightarrow \mathbf{data} \coloneqq exp$ | Find $p$ or fail | Update $key$ |
| $d \coloneqq p \rightarrow \mathbf{data}$ | Find $p$ or fail | Update $d$ |
| $\mathbf{new}\ p$ | Move to the first inactive child or fail | $\langle p \coloneqq \mathbf{here} \rangle$ and set $active$ |
| $\mathbf{free}\ p$ | Find $p$ or fail | Reset $active$ |

Table 1. Summary of the encoding of the walking statements.

**Encoding of new $p$.** Let $j$ be the smallest index in $[k+1, k+m]$ such that $active\_child_j$ is *false*. If no such index exists, we push a frame on $\bar{t}$ with the symbolic instruction OOM, representing an out-of-memory error. Otherwise, the lace moves to the $j^{\text{th}}$ child of $\bar{t}$ and pushes a frame $f$ there with $f.pc = succ(\bar{f}.pc)$, $f.instr = \langle p \coloneqq \mathbf{here} \rangle$, $f.active = true$, and $f.isnil_p = false$.

**Encoding of free $p$.** The lace moves to the node $u$ pointed by $p$ and pushes there a frame $f$ with $active = false$ and an updated $pc$. In the new frame, $isnil_q$ is set to *true* for every pointer $q$ that currently points at $u$, including $p$. To find such pointers, let $(u, i)$ be the id of $f$. Then, $q$ points at $u$ if the log of $u$ contains another frame $(u, j)$ such that: $j < i$, $(u, j)$ contains $\langle q \coloneqq \mathbf{here} \rangle$, and the $upd_q$ flag is *false* in all frames from $(u, j+1)$ to $(u, i)$.

Table 1 summarizes the previous cases and includes the encodings for the statements $p \rightarrow \mathbf{data} \coloneqq exp$ and $d \coloneqq p \rightarrow \mathbf{data}$, which are simple variations of the ones described above.

**Encoding Boolean Conditions and Control-flow Statements.** Conditionals and loops can be local or walking instructions, depending on the Boolean condition. Data conditions are evaluated locally by using the values of local variables stored in the current frame $\bar{f}$. For heap conditions, we may traverse the lace. Here, we only consider conditions of the form $p = q$, as we can simulate other heap conditions with extra variables. For example, we replace $p \rightarrow pfield = q$ with $r = q$ by introducing a new pointer variable $r$ and assigning it the value of $p \rightarrow pfield$.

We handle conditions of the form $p = q$ as follows. We first attempt to resolve the heap condition using $isnil$ flags. If both pointers have their $isnil$ flags set to true, the condition is *true*; if the flags differ, the condition evaluates to false. if the local resolution is inconclusive, we rewind the lace starting from $\bar{f}$ until finding an assignment to either $p$ or $q$. For instance, encountering $\langle p \coloneqq \mathbf{here} \rangle$ within frame $(u, i)$, we then search through the frames of $u$'s log for largest index $j < i$ where frame $(u, j)$ contains $\langle q \coloneqq \mathbf{here} \rangle$. If no such index exists, the condition is *false*. If found, the condition holds if and only if $q$ was not updated from $(u, j)$, checked using the $upd_q$ flags: $\neg \bigvee_{l \in [j+1, i]} upd_q^l$. Based on this evaluation, we push a new frame where we update the program counter $pc$ based on the condition outcome. If we first encounter $\langle q \coloneqq \mathbf{here} \rangle$, the process is symmetrical.

In the following, for a label $\sigma$ and $i \in \mathbb{N}$, we denote by $\sigma^{<i}$ (resp., $\sigma^{\leq i}$) the label obtained by setting *avail* to true in all frames of $\sigma$ with indices greater than or equal to $i$ (greater than $i$, resp.).

The following lemma states that each new frame in a knitted-tree depends only on *local* information from the neighboring nodes, as detailed in the construction. Only the *next* field remains unconstrained until the next frame is known. Hence, we write $f_1 \equiv f_2$ to signify that the frames $f_1$ and $f_2$ are equal except for their *next* field.

LEMMA 4.3 (LOCALITY). *There exist functions Up, Down, Internal with the following properties. Let $\sigma, \tau_1, \ldots, \tau_{k+m}$ be the logs of a node $u$ and of its children in a knitted-tree prefix.*

- *For all steps $(u.j, b) \rightarrow_{next} (u, a)$ in the lace, it holds $\sigma^a \equiv Up(\tau_j^{\leq b}, j, \sigma^{<a})$.*
- *For all steps $(u, a) \rightarrow_{next} (u.j, b)$ in the lace, it holds $\tau_j^b \equiv Down(\sigma^{\leq a}, j, \tau_j^{<b})$.*
- *For all steps $(u, a) \rightarrow_{next} (u, a+1)$ in the lace, it holds $\sigma^{a+1} \equiv Internal(\sigma^{\leq a})$.*

*On the choice of parameters $m$ and $n$.* Parameter $m$ corresponds to the number of allocations made by the program. Programs that do not create new nodes can be analyzed with $m = 0$, while $m = 1$ is adequate for programs that insert a single new node. Parameter $n$ relates to the number of passes and instructions executed per node. Although some terminating programs may require unbounded labels, moderate $n$ values typically suffice for common tree-like algorithms.

## 5 Properties of knitted-trees

*Prefix of a knitted-tree.* An $(m, n)$-knitted-tree is defined as a knitted-tree associated with an execution $\pi$ that belongs to $kt(\pi, m, n)$. A knitted-tree *prefix* of a knitted-tree $\mathcal{K}$ is any data tree derived from $\mathcal{K}$ by truncating its lace at some frame $f$ (i.e., setting the flag $f'.avail$ to *true* for all frames $f'$ that follow $f$ in the lace), and unconstraining $f.next$ so that it can take any value.

### 5.1 Compositionality

**Definition of the predicate formula** *consistent_child*$(\tau, j, \sigma)$. Based on the functions *Up* and *Down* from Lemma 4.3, we define the formula *consistent_child*$(\tau, j, \sigma)$ using the underlying data theory $\mathcal{D}$. This predicate assumes that $\sigma$ and $\tau$ are logs belonging to two possibly different knitted-trees and ensures that they occur in the *same* knitted-tree as the logs of a node and its $j^{\text{th}}$ child. To ensure this property, it checks that consecutive frames $(f_1, f_2)$ connected by the attributes *next* and *prev*, with one frame belonging to $\tau$ and the other frame belonging to $\sigma$, adhere to the functions *Up* and *Down* accordingly. A precise definition of *consistent_child* can be found in Appendix C.4.

The next lemma directly follows from the definitions of *consistent_child* and knitted-tree prefix, affirming that *consistent_child* holds for all pairs of parent-child logs of a knitted-tree.

LEMMA 5.1. *For all labels $\sigma, \tau \in L(\mathcal{S}_{\mathcal{K}})$ and indices $j \in [k + m]$, if there exists a knitted-tree prefix where $\sigma$ and $\tau$ are the logs of a node and its $j^{th}$ child respectively, then consistent_child$(\tau, j, \sigma)$ holds.*

The following lemma establishes the key property of *consistent_child* relevant to our verification approach, enabling the composition of knitted-trees by combining subtrees from different instances.

LEMMA 5.2 (COMPOSITIONALITY). *For $i = 1, 2$, let $\sigma_i \in L(\mathcal{S}_{\mathcal{K}})$ be the log of a node $t_i$ in an $(m, n)$-knitted-tree prefix $\mathcal{K}_i$. If consistent_child$(\sigma_2, j, \sigma_1)$ holds true for some $j \in [k + m]$, then there exists an $(m, n)$-knitted-tree prefix $\mathcal{K}$ where $\sigma_1$ is the log of a node and $\sigma_2$ is the log of its $j^{th}$ child. Moreover, $\mathcal{K}$ can be obtained by replacing the subtree rooted at the $j^{th}$ child of $t_1$ in $\mathcal{K}_1$ with the subtree rooted at $t_2$ in $\mathcal{K}_2$.*

PROOF. Let $\mathcal{K}$ be the data tree obtained by replacing the subtree rooted at the $j^{th}$ child of $t_1$ in $\mathcal{K}_1$ with the subtree rooted at $t_2$ in $\mathcal{K}_2$. We prove that $\mathcal{K}$ is a knitted-tree prefix by induction on the number of pairs $(a, b)$ where frames $\sigma_1^a$ and $\sigma_2^b$ are adjacent in the lace, each such pair representing an interaction between the parent's and the child's labels.

When the above number is zero, there are no interactions between $t_1$ and its $j^{th}$ child in $\mathcal{K}_1$. Let $\pi_1$ be an execution s.t. $\mathcal{K}_1$ is a prefix of a knitted-tree representing $\pi_1$. It is direct to show that $\mathcal{K}$ is a knitted-tree prefix for an execution $\pi$ following the same steps as $\pi_1$, starting with the input tree of $\mathcal{K}$. Since $\pi_1$ never visits the $j^{\text{th}}$ child of $t_1$, and this subtree is the only difference between $\mathcal{K}_1$ and $\mathcal{K}$, $\pi$ is a valid execution of $P$.

For the inductive case, consider the last interaction $(a, b)$ between $\sigma_1^a$ and $\sigma_2^b$. First, assume that such interaction is a step *up* from the frame $\sigma_2^b$ to $\sigma_1^a$. Let $\mathcal{K}_1'$ be derived from $\mathcal{K}_1$ by truncating its lace to end just before $\sigma_1^a$. Clearly, $\mathcal{K}_1'$ is still a knitted-tree prefix, and the modified label $\sigma_1'$ of $t_1$ is obtained from $\sigma_1$ by removing the frames with index at least $a$, by setting their *avail* flags to true. We can now apply the inductive hypothesis to the labels $\sigma_1'$ and $\sigma_2$, because we have removed one interaction between them. Hence, we can assume that there is a single knitted-tree prefix $\mathcal{K}'$ containing both labels as the logs of a parent and its $j^{\text{th}}$ child. Next, we obtain the desired knitted-tree prefix $\mathcal{K}$ from $\mathcal{K}'$ by reintroducing the sequence of frames removed from $\mathcal{K}_1$. We need to prove that adding those frames respects all rules of knitted-trees. The correctness of the first added frame, $\sigma_1^a$, is ensured by *consistent_child*$(\sigma_2, j, \sigma_1)$, because it applies the function *Up* to all upward interactions between $\sigma_2$ and $\sigma_1$. In turn, Lemma 4.3 ensures that adhering to that function is sufficient to establish the correctness of the next frame. Subsequent frames can be reintroduced due to unchanged surroundings remain. Lemma 4.3 ensures that no other information is relevant.

The other case to prove is when the last interaction is a step *down* from the parent's frame $\sigma_1^a$ to the child's frame $\sigma_2^b$. Define $\sigma_2'$ as the label obtained from $\sigma_2$ with the frames of indices $b$ and above removed. Similar to the previous case, we apply the inductive hypothesis to the shortened label $\sigma_2'$ and its shortened knitted-tree prefix $\mathcal{K}_2'$, resulting in a knitted-tree prefix $\mathcal{K}'$. We then reintroduce the frames removed from $\mathcal{K}_2$ into $\mathcal{K}'$. The correctness of the first reintroduced frame is ensured by *consistent_child* checking the function *Down* on every downward interaction. The subsequent reintroduced frames are still valid because there are no steps returning to the parent of $t_2$, and their surroundings remain unchanged from $\mathcal{K}_2$. □

## 5.2 Relations with Program Executions

*Retrieving the heap from its knitted-tree representation.* We select the nodes where *active* is *true* in their top frame. For a node $u$ and pointer field *pfield* (assumed to be the $j^{\text{th}}$ field in *PF*):

- If *pfield* has never been updated, $u.pfield$ points to the $j^{\text{th}}$ child of $u$, if $u.j$ exists and is active in the knitted-tree; otherwise, $u.pfield$ points to **nil**.
- If the last update in $u$'s log is $\langle pfield := \textbf{nil} \rangle$, then pointer field is also **nil**.
- Otherwise, the last update is $\langle pfield := p \rangle$ for some pointer variable $p$ (not **nil**). We navigate the lace backward using *find_ptr* until finding a frame that reports $\langle p := \textbf{here} \rangle$, say in the log of node $v$, and set $u.pfield$ to $v$.

We denote by *heap*$(\mathcal{K})$ the heap corresponding to the knitted-tree (prefix) $\mathcal{K}$, as described above.

From a knitted-tree, we can find the final value of each pointer variable. We denote by *pv*$(\mathcal{K})$ the map that assigns each $p \in PV_P$ to the node of *heap*$(\mathcal{K})$ pointed to by $p$ at the end of the lace of $\mathcal{K}$.

*Exit status of a knitted-tree.* To distinguish how knitted-trees terminate, we introduce the notion of *exit statuses* for individual frames and for the entire knitted-tree. Each frame $f$ of a knitted-tree is assigned one of five statuses in *ExStatus* = $\{\textbf{C}, \textbf{E}, \textbf{O}, \textbf{M}, \textbf{N}\}$, with the following meanings:

- **C**lean exit: The program counter (*pc*) of $f$ points to an **exit** instruction;
- Runtime **E**rror: $f.instr = \textsc{err}$, indicating a null-pointer dereference;
- Label **O**verflow: the index of $f$ in its label is $n + 1$, indicating log overflow;
- Out of **M**emory: $f.instr = \textsc{oom}$, indicating a failed attempt to allocate a node with the **new** statement due to the absence of inactive nodes.
- **N**one: Indicates that frame $f$ is not a terminal frame in all other cases.

A frame $f$ is *terminal* if its status is not $\mathbf{None}$. Indeed, the exit statuses different from $\mathbf{N}$ terminate the lace, hence only the last frame in the lace may have an exit status different from $\mathbf{N}$. The *exit status* of a knitted-tree is the status of the last frame in its lace.

The following lemmas relate the exit status of a knitted-tree to the execution it was derived from.

LEMMA 5.3. *For all executions $\pi$ and knitted-trees $\mathcal{K} \in kt(\pi, m, n)$, if $\mathcal{K}$'s exit status is not in $\{\mathbf{O}, \mathbf{M}\}$, then the last configuration of $\pi$ is $(heap(\mathcal{K}), pv(\mathcal{K}), f.d, f.pc)$, where $f$ is the last frame of $\mathcal{K}$'s lace.*

PROOF. Let $\mathcal{K} = (K, \mu)$. The proof is by induction on the length of $\pi$. The base case is when $\pi$ consists only of the initial configuration. The thesis follows from the definition of the backbone $K$ and the base case of the labeling function $\mu$.

For the inductive case, consider the prefix $\overline{\pi}$ of $\pi$ with one fewer step. We remove from $\mathcal{K}$ the sequence of frames encoding the last step of $\pi$, by setting the *avail* fields of those frames to *true*. The resulting data tree is a knitted-tree $\overline{\mathcal{K}} \in kt(\overline{\pi}, m, n)$. By inductive hypothesis, the last configuration of $\overline{\pi}$ can be recovered from $\overline{\mathcal{K}}$. By inspecting the encoding of the various types of statements, as described in §4.3, we can prove that the encoding in $\mathcal{K}$ of the last step in $\pi$ faithfully preserves the information required to reconstruct the last configuration in $\pi$. □

LEMMA 5.4. *Let $\mathcal{K} \in kt(\pi, m, n)$ for some execution $\pi$ and parameters $m$ and $n$. If the exit status of $\mathcal{K}$ is $\mathbf{E}$, then $\pi$ ends in an error configuration. Conversely, if $\pi$ ends in an error configuration, then the exit status of $\mathcal{K}$ is in $\{\mathbf{E}, \mathbf{O}, \mathbf{M}\}$.*

PROOF. To prove the first statement, assume that the exit status of $\mathcal{K}$ is $\mathbf{E}$, occurring when the last frame $f$ in $\mathcal{K}$'s lace has the symbolic instruction ERR. By Lemma 5.3, the last configuration of $\pi$ can be recovered from $\mathcal{K}$ as the tuple $c = (heap(\mathcal{K}), pv(\mathcal{K}), f.d, f.pc)$. By definition of knitted-trees, the symbolic instruction ERR arise only from a null pointer dereference, so $c$ is an error configuration.

For the second statement, assume that $\pi$ ends in an error configuration and the exit status of $\mathcal{K}$ is not $\mathbf{O}$ and $\mathbf{M}$. We show that the exit status of $\mathcal{K}$ is $\mathbf{E}$. By Lemma 5.3, the last configuration of $\pi$ can be recovered from $\mathcal{K}$ as $(heap(\mathcal{K}), pv(\mathcal{K}), f.d, f.pc)$, where $f$ is the last frame of the lace of $\mathcal{K}$. Since $\pi$ ends in an error configuration, the program counter at $f$ points to an instruction causing a null pointer dereference, leading to the symbolic instruction ERR and hence to the exit status $\mathbf{E}$. □

LEMMA 5.5. *For all $m, n$, the $(m, n)$-knitted-trees of infinite executions have exit status $\mathbf{O}$.*

PROOF. Each step of an execution adds at least one frame to the lace of its knitted-tree. Since the total number of nodes and frames in a knitted-tree is finite, the lace will eventually occupy a frame with index $n + 1$ in the log hosting that frame, resulting in the exit status $\mathbf{O}$. □

## 6 Reasoning about Knitted-trees through Constrained Horn Clauses

In this section, we define a system of CHCs whose minimal model characterizes the set of all logs of *any* $(m, n)$-knitted-tree prefix of a given program.

### 6.1 The CHC System

For a program $P$ and integer parameters $k$, $m$, and $n$, we introduce a CHC system called $C_{kt}(P, k, m, n)$. This system uses a single uninterpreted relation symbol, $\mathbf{Lab}(\sigma)$, with $\sigma$ matching the data signature $\mathcal{S}_{\mathcal{K}}$ of knitted-trees. The design of $C_{kt}(P, k, m, n)$ guarantees the following property:

THEOREM 6.1. *In the minimal model of the CHC system $C_{kt}(P, k, m, n)$, $\mathbf{Lab}(\sigma)$ is true for a label $\sigma$ if and only if there exists an $(m, n)$-knitted-tree prefix $\mathcal{K}$ of $P$ such that $\sigma$ is the label of a node in $\mathcal{K}$.*

| | | |
|---|---|---|
| **(I)** | $\mathbf{Lab}(\sigma) \leftarrow len(\sigma, 1) \wedge first\_frame(\sigma^1)$ | *Initializing non-root nodes* |
| **(II)** | $\mathbf{Lab}(\sigma) \leftarrow len(\sigma, 2) \wedge start(\sigma)$ | *Initializing the root node* |
| **(III)** | $\mathbf{Lab}(\sigma) \leftarrow len(\sigma, i) \wedge \mathbf{Lab}(\sigma^{<i}) \wedge \Psi_{Internal}(\sigma^{<i}, \sigma^i)$ | *An internal step* |
| **(IV)** | $\mathbf{Lab}(\sigma) \leftarrow \ len(\sigma, i) \wedge \mathbf{Lab}(\sigma^{<i}) \wedge \mathbf{Lab}(\tau)$ | *A step from the $j^{th}$ child to its parent* |
| | $\qquad \wedge\ consistent\_child(\tau, j, \sigma^{<i})$ | |
| | $\qquad \wedge\ \Psi_{Up}(\tau, j, \sigma^{<i}, \sigma^i)$ | |
| **(V)** | $\mathbf{Lab}(\tau) \leftarrow len(\tau, i) \wedge \mathbf{Lab}(\sigma) \wedge \mathbf{Lab}(\tau^{<i})$ | *A step from a node to its $j^{th}$ child* |
| | $\qquad \wedge\ consistent\_child(\tau^{<i}, j, \sigma)$ | |
| | $\qquad \wedge\ \Psi_{Down}(\sigma, j, \tau^{<i}, \tau^i)$ | |
| **(VI)** | $\bot \leftarrow \mathbf{Lab}(\sigma) \wedge label\_exit(\sigma, Ex)$ | *The lace ends with exit status in Ex* |

Fig. 5. CHCs (I)-(V) form the CHC system $C_{kt}(P, k, m, n)$, while the CHC system $C_{ex}(\mathcal{I})$ includes all the CHCs in the figure. Here, $i \in [2, n]$ and $j \in [k + m]$.

Our approach to defining CHCs in $C_{kt}(P, k, m, n)$ builds on the rules for constructing knitted-trees from §4.3, with a twist. In §4.3, we extend a partial knitted-tree by adding frames to existing logs. However, building a partial knitted-tree is impractical due to our predicate's nature, and we must reason about labels from *all* possible knitted-trees. To address this, we use the compositionality lemma (Lemma 5.2), which states that two consistent labels imply a knitted-tree where those labels are logs of an internal node and one of its children. Additionally, we use the locality lemma (Lemma 4.3) to extend the lace involving these nodes. This lemma entails that constructing a knitted-tree involves adding frames to node logs so that any two consecutive frames belong to the same node or to neighboring backbone nodes. We use this property in the CHC system, employing independent CHCs to simulate adding a single frame. Specifically, we use the functions *Up* for upward movement, *Down* for downward movement, and *Internal* for expanding the lace within the same log. To use them in the CHC system, we introduce three new predicates to constrain logs according to these functions: $\Psi_{Down}(\sigma, j, \tau_j, f)$, $\Psi_{Up}(\tau_j, j, \sigma, f)$, and $\Psi_{Internal}(\sigma, f)$, where $f$ represents the frame output by the corresponding function. See the Appendix C.4 for details.

Fig. 5 details the CHCs of $C_{kt}(P, k, m, n)$. While describing each CHC, we establish the "only if" direction of Theorem 6.1, by induction on the number of CHC applications needed to insert $\sigma$ into the minimal interpretation of **Lab**. The proof of the "if" direction follows the CHC descriptions.

Before detailing the CHCs, we introduce some notation. Let $\sigma \in L(\mathcal{S}_\mathcal{K})$ and $i \in \mathbb{N}$. The formula $len(\sigma, i)$ is true if all frames with indices in $[i]$ are unavailable and all other frames are available, i.e., $len(\sigma, i) \stackrel{\text{def}}{=} \neg\sigma^i.avail \wedge \bigwedge_{j=i+1}^{n+1} \sigma^j.avail$. With an abuse of notation, we write $\sigma^{<i}$ in a CHC as a shorthand for a fresh variable $\theta$, together with the conjunct $\bigwedge_{\ell \in [i-1]}(\theta^\ell = \sigma^\ell) \wedge \bigwedge_{\ell \in [i, n+1]} \theta^\ell.avail$.

**CHC I** and **CHC II** ensure that **Lab** includes the labels of all knitted-trees representing an execution of $P$ of length of 0. These labels form the base case for the proof of the "only if" direction of Theorem 6.1, as both CHCs are facts. CHC I defines the labels of non-root nodes where all frames except the first are available, and the fields *active* and *active_child* are consistent (see Appendix C.1). CHC II specifies the root node labels, constraining the first two frames to be unavailable, with $start(\sigma)$ specifying the second frame of $\sigma$ according to the base case of knitted-tree labels (see §4.3.1 and Appendix C.2).

The remaining CHCs extend each node's log one frame at a time, following the inductive case of the knitted-tree label definition. To prove the "only if" direction of Theorem 6.1, we inductively assume that the labels in the body of each rule satisfy the claim, i.e., they label a node in a $(m, n)$-knitted-tree prefix. We now show that the label in the rule's head also satisfies the claim.

**CHC III** addresses cases where $\sigma^i$ follows $\sigma^{i-1}$ in the lace, aka an internal step. In the CHC, $\mathbf{Lab}(\sigma^{<i})$ ensures that $\sigma^{<i}$ labels a node in an $(m, n)$-knitted-tree prefix. The formula $\Psi_{Internal}(\sigma^{<i}, \sigma^i)$ constrains the next frame $\sigma^i$ to encode the next internal step, as asserted in Lemma 4.3.

**CHC IV** addresses cases where the lace extends with a new frame pushed to the parent of the previous frame, typically during a rewind phase of a walking statement. The predicate $consistent\_child$ ensures that the labels $\sigma^{<i}$ and $\tau$ belong to the same knitted-tree prefix, as the log of a parent and its $j^{\text{th}}$ child (Lemma 5.2). Then, the predicate $\Psi_{Up}$ extends the lace by adding an extra frame to $\sigma^{<i}$, following the topmost frame of $\tau$.

**CHC V** addresses the opposite scenario of CHC IV, where the last step of the current lace prefix goes from the parent to its $j^{\text{th}}$ child. This rule, thanks to $\Psi_{Down}$, ensures that $\tau_j$ correctly extends $\tau_j^{<i}$ with an additional frame representing the latest step in the lace.

PROOF. ("if" direction of Theorem 6.1). Let $t$ be a node in $\mathcal{K}$ with log $\sigma$. We proceed by induction on the length $\ell$ of the lace of $\mathcal{K}$.

*Base case* ($\ell = 1$): There is only one frame in the lace at the root of $\mathcal{K}$. Here, either CHC I or CHC II inserts $\sigma$ into $\mathbf{Lab}$, depending on whether 1 or 2 frames are unavailable in $\sigma$.

*Inductive step* ($\ell > 1$): Consider $\mathcal{K}_{\ell-1}$, the prefix of $\mathcal{K}$ with the last frame of the lace removed. By inductive hypothesis, the logs of $\mathcal{K}_{\ell-1}$ belong to $\mathbf{Lab}$. If $\sigma$ is a log of $\mathcal{K}_{\ell-1}$, the claim holds. Otherwise, $\sigma$ contains the last frame in the lace of $\mathcal{K}$, meaning that the logs of $t$'s neighboring nodes (its parent and children) are unchanged from $\mathcal{K}_{\ell-1}$. Thus, by inductive hypothesis those logs belong to $\mathbf{Lab}$. Depending on the direction taken in the last step, $\sigma$ is added to $\mathbf{Lab}$ by one of CHC III to CHC V, as detailed in the above description of the individual CHCs. □

## 6.2 The Exit Status Problem

In this section, we propose a method to determine if a given program can lead to runtime memory safety error via an execution that can be represented by an $(m, n)$-knitted-tree. We achieve this by solving a CHC system. If the system is unsatisfiable, it indicates the existence of an execution leading to a runtime error. Formally, the decision problem addressed in this section is the following.

PROBLEM 1 (EXIT STATUS PROBLEM). *An instance $\mathcal{I}$ of the **exit status problem** is a tuple $(P, k, m, n, Ex)$, where $P$ is a program that operates on $k$-ary $\mathcal{S}^{\Sigma}$-trees with $\mathcal{S}^{\Sigma} = \{key : \mathcal{D}\}$, $m, n \in \mathbb{N}$, and $Ex$ is a set of exit statuses excluding $\mathbf{N}$. The exit status problem asks whether there exists an $(m, n)$-knitted-tree of $P$ whose exit status belongs to $Ex$.*

We solve the exit status problem via the CHC system $\mathcal{C}_{\text{ex}}(\mathcal{I})$: the exit status problem is positive iff $\mathcal{C}_{\text{ex}}(\mathcal{I})$ is unsatisfiable. Fig. 5 shows the CHCs of $\mathcal{C}_{\text{ex}}(\mathcal{I})$. The CHCs of $\mathcal{C}_{\text{ex}}(\mathcal{I})$ are:

- all CHCs of $\mathcal{C}_{\text{kt}}(P, k, m, n)$ that are essential for establishing Theorem 6.1.
- a unique query, **CHC VI**, to check for the existence of a log of a knitted-tree (prefix) corresponding to a program execution with an exit status in $Ex$.

The following theorem, which is the main result of this section, directly follows from Theorem 6.1 and the definition of CHC VI.

**Theorem 6.2.** *Let $\mathcal{I}$ be an instance of the exit-status problem. Then, $\mathcal{I}$ admits a positive answer if and only if the CHC system $C_{\mathrm{ex}}(\mathcal{I})$ is unsatisfiable.*

**Proof.** Let $\mathcal{I} = (P, k, m, n, Ex)$. We prove the theorem by establishing the equivalent statement: $\mathcal{I}$ *admits a negative answer if and only if the CHC system $C_{\mathrm{ex}}(\mathcal{I})$ is satisfiable.*

For the "only if" direction, if $\mathcal{I}$ admits a negative answer, no $(m, n)$-knitted-tree of $P$ has an exit status in $Ex$. Consider the interpretation for **Lab** containing all and only the node labels of all $(m, n)$-knitted-tree prefixes of $P$. By Theorem 6.1, this interpretation is in fact the minimal model of $C_{\mathrm{kt}}(P, k, m, n)$, and as such it satisfies the CHCs I-V. Moreover, since none of those labels has an exit status in $Ex$, this interpretation also satisfies the query VI, showing that $C_{\mathrm{ex}}(\mathcal{I})$ is satisfiable.

For the "if" direction, if $C_{\mathrm{ex}}(\mathcal{I})$ is satisfiable, the interpretation of **Lab** in any model of $C_{\mathrm{ex}}(\mathcal{I})$ has the property that no label in **Lab** has an exit status in $Ex$ (CHC VI). Since $C_{\mathrm{ex}}(\mathcal{I})$ implies $C_{\mathrm{kt}}(P, k, m, n)$, by Theorem 6.1 such interpretation of **Lab** contains at least all the logs of the $(m, n)$-knitted-tree (prefixes) of $P$. We conclude that the answer to the exit status problem is negative. □

## 7 CHC-based Verification of programs complying with precondition

This section focuses on solving the exit-status problem with preconditions, investigating how to handle and reason about them within the CHC systems introduced in §6.

We use SDTAs (introduced in §2.1) to specify preconditions. We denote by $\mathcal{A}_{\mathrm{pre}} = (\mathcal{S}^{\Sigma}, \mathcal{S}^{Q}, \psi, \psi^{F})$ the SDTA defining the set of feasible input data trees for $P$. $\mathcal{A}_{\mathrm{pre}}$ operates on $k$-ary data trees over the data signature $\mathcal{S}^{\Sigma} = \{key : \mathcal{D}\}$. Here, $\psi(q_1, \ldots, q_k, key, q)$ is the transition constraint, and $\psi^{F}(q)$ defines the set of its final states.

Similarly to the function $heap(\mathcal{K})$ that extracts the output heap from a knitted-tree, we define $in(\mathcal{K})$ to extract the input data tree from a knitted-tree $\mathcal{K}$. For $\mathcal{K} = (K, \mu)$, $in(\mathcal{K})$ is the $\mathcal{S}^{\Sigma}$-tree $(T, \lambda)$, where $T$ is empty if the first frame of the root of $K$ is inactive. Otherwise, $T$ contains all internal nodes of $K$, and for each node $t \in T$, $\lambda(t).key = \mu(t).key^1$.

We refine the exit status problem to comply with preconditions as follows.

**Problem 2 (Exit Status Problem with Precondition).** *An instance of the* **exit status problem with precondition** *is a pair $(\mathcal{I}, \mathcal{A}_{\mathrm{pre}})$, where $\mathcal{I} = (P, k, m, n, Ex)$ is an instance of the exit status problem and $\mathcal{A}_{\mathrm{pre}}$ is an SDTA on $k$-ary $\mathcal{S}^{\Sigma}$-trees. We say that $(\mathcal{I}, \mathcal{A}_{\mathrm{pre}})$ admits a positive answer if and only if there exists an $(m, n)$-knitted-tree $\mathcal{K}$ of $P$ such that: $\mathcal{K}$ satisfies the exit status problem $\mathcal{I}$ and $in(\mathcal{K})$ belongs to $L(\mathcal{A}_{\mathrm{pre}})$.*

We reduce the exit status problem with precondition to CHC satisfiability. For an instance $(\mathcal{I}, \mathcal{A}_{\mathrm{pre}})$, we define the system $C_{\mathrm{pre}}(\mathcal{I}, \mathcal{A}_{\mathrm{pre}})$ as shown in Fig. 6. Besides the CHCs in the figure, it includes all $C_{\mathrm{kt}}(P, k, m, n)$'s clauses. We introduce a fresh uninterpreted relation symbol $\mathbf{Pre}(\sigma, q, e)$, where $\sigma$ aligns with $\mathcal{S}_{\mathcal{K}}$, $q$ is of type $\mathcal{S}^{Q}$, and $e$ is Boolean. $C_{\mathrm{pre}}(\mathcal{I}, \mathcal{A}_{\mathrm{pre}})$ is defined to ensure the following property.

**Theorem 7.1.** *Consider the minimal model of* **Pre** *and* **Lab** *that satisfies all the clauses of the CHC system $C_{\mathrm{pre}}(\mathcal{I}, \mathcal{A}_{\mathrm{pre}})$ except for its query. It holds that:* $\mathbf{Pre}(\sigma, q, e)$ *holds true if and only if there exist:*

*(a) a prefix $\mathcal{K}$ of an $(m, n)$-knitted-tree of $P$;*

*(b) a node $t$ in $\mathcal{K}$;*

*(c) a run $\rho$ of $\mathcal{A}_{\mathrm{pre}}$ on the subtree of $in(\mathcal{K})$ rooted at $t$;*

*such that all the following conditions hold:*

*(1) $\sigma$ is the log of $t$ in $\mathcal{K}$;*

*(2) $q$ is* **nil** *if $t$ is an auxiliary node of $\mathcal{K}$, and $\rho(t)$ otherwise;*

*(3) $e$ is true if there exists a node in the subtree of $\mathcal{K}$ rooted at $t$ whose top frame has an exit status belonging to $Ex$.*

$$\textbf{(i)} \quad \textbf{Pre}(\sigma, q, e) \leftarrow \textbf{Lab}(\sigma) \qquad\qquad\qquad\qquad \textit{Labeling the leaves}$$
$$\wedge \neg(\sigma.active^1) \wedge \neg start(\sigma) \wedge q = \textbf{nil}$$
$$\wedge \left( e \leftrightarrow label\_exit(\sigma, Ex) \right)$$

$$\textbf{(ii)} \quad \textbf{Pre}(\sigma, q, e) \leftarrow \textbf{Lab}(\sigma) \wedge \textbf{Pre}(\tau_1, q_1, e_1) \wedge \ldots \wedge \textbf{Pre}(\tau_{k+m}, q_{k+m}, e_{k+m})$$
$$\wedge \bigwedge_{j \in [k+m]} consistent\_child(\tau_j, j, \sigma) \qquad \textit{Labeling internal nodes}$$
$$\wedge \Big( \left( \sigma.active^1 \wedge \psi(q_1, \ldots, q_k, \sigma.key^1, q) \right)$$
$$\vee \left( \neg \sigma.active^1 \wedge q = \textbf{nil} \right) \qquad\qquad \Big)$$
$$\wedge \left( e \leftrightarrow \left( e_1 \vee \ldots \vee e_{k+m} \vee label\_exit(\sigma, Ex) \right) \right)$$

$$\textbf{(iii)} \qquad \bot \leftarrow \textbf{Pre}(\sigma, q, e) \wedge start(\sigma) \wedge e \wedge \psi^F(q) \qquad \textit{Checking the root}$$

Fig. 6. The CHC system $C_{\text{pre}}(I, \mathcal{A}_{\text{pre}})$ includes the CHCs in this figure and the CHCs I-V from Fig. 5.

We now describe the CHCs shown in Fig. 6. Simultaneously, we establish the "only if" direction of Theorem 7.1 through induction on the number of CHC applications required to insert the triple into the minimal interpretation of **Pre**. The proof of the "if" direction follows the CHC descriptions.

Observe that in the context of $C_{\text{pre}}(I, \mathcal{A}_{\text{pre}})$, **Lab** is independent of **Pre**, so the CHCs shown in Fig. 5 retain their original meaning and properties. Specifically, Theorem 6.1 ensures that all labels satisfying **Lab** occur as logs in some prefix of an $(m, n)$-knitted-tree of the program $P$.

**CHC (i)** inserts the logs of the knitted-tree leaves into **Pre**, with extra information. This serves as the base case in the inductive proof of Theorem 7.1. Leaves are nodes that are initially inactive and do not satisfy the predicate $start$. In such cases, **nil** is inserted as the second component of the triple. The quantifier-free predicate $label\_exit(\sigma, Ex)$ sets the flag $e$, evaluating to $true$ if and only if the top frame of $\sigma$ corresponds to the lace tail and its exit status belongs to $Ex$ (see Appendix C.3 for details). The constraint $\neg start(\sigma)$ excludes the case when the input tree is empty and $\sigma$ is the root label of the knitted-tree, which is covered by CHC (ii).

**CHC (ii)** propagates the state of $\mathcal{A}_{\text{pre}}$ bottom-up. Inspecting its body, $\textbf{Lab}(\sigma)$ ensures that $\sigma$ is a log in a knitted-tree prefix, due to Theorem 6.1. By inductive hypothesis, for all $j \in [k + m]$ the rest of the first line ensures the existence of a subtree $\mathcal{K}_j$ of a knitted-tree, whose root is labeled with $\tau_j$. Moreover, $q_j$ is the state assigned to the root of $\mathcal{K}_j$ by a run of $\mathcal{A}_{\text{pre}}$ on $in(\mathcal{K}_j)$, and $e$ is $true$ if $\mathcal{K}_j$ contains a log whose exit status belongs to $Ex$.
The second line ensures that the labels under consideration are the logs of a node and its $k + m$ children in the *same* knitted-tree prefix. This follows from repeatedly applying Lemma 5.2 to each label $\tau_j$. The third line uses $\mathcal{A}_{\text{pre}}$'s transition constraint $\psi$ to determine the new state $q$ in the parent node, excluding the special case when $\sigma$ labels the root of a knitted-tree with an empty input tree. In that case, the next line assigns state **nil** to the root. Finally, the last line sets the parent's exit status flag $e$ to $true$ if any child has the same flag set to $true$ or if $label\_exit$ evaluates to $true$ on $\sigma$.

**CHC (iii)** is the only query in the CHC system. It makes the system unsatisfiable if, during the simulation of $\mathcal{A}_{\text{pre}}$, the root of any knitted-tree (prefix) is labeled with a final state of $\mathcal{A}_{\text{pre}}$ and the parameter $e$ is *true*. This indicates that a knitted-tree satisfying the precondition contains an end-of-lace frame with an exit status in *Ex*.

It is worth clarifying how an empty input tree is handled. The precondition, $\mathscr{A}_{\mathrm{pre}}$, may or may not accept an empty input tree. The system of CHCs handles both cases correctly: for an empty input tree, we get a knitted-tree with two levels of auxiliary nodes (*active*[1] set to *false*). Thus, CHC (i) labels these nodes with $q = \mathbf{nil}$, CHC (ii) puts $\mathbf{nil}$ also on the root, and CHC (iii) evaluates the acceptance constraint $\psi^F(\mathbf{nil})$, reflecting whether $\mathscr{A}_{\mathrm{pre}}$ accepts the empty tree.

Proof. ("*if*" *direction of Theorem 7.1*). Assume that $\mathcal{K}$, $t$, and $\rho$ satisfy the assumptions of the theorem. Let $\mathbb{I}(\mathbf{Pre})$ (resp., $\mathbb{I}(\mathbf{Lab})$) be an interpretation of $\mathbf{Pre}$ (resp., $\mathbf{Lab}$) corresponding to the minimal model of $C_{\mathrm{pre}}(\mathcal{I}, \mathscr{A}_{\mathrm{pre}})$. We prove by structural induction that $\mathbb{I}(\mathbf{Pre})(\sigma, q, e)$ is true, where the three components comply with items (1)-(3) of Theorem 7.1. Note that Theorem 6.1 ensures that $\mathbb{I}(\mathbf{Lab})(\sigma)$ holds.

The base case arises when $t$ is a leaf of $\mathcal{K}$. Then, CHC (i) inserts the appropriate triple into $\mathbb{I}(\mathbf{Pre})$.

The inductive step occurs when $t$ is an internal node of $\mathcal{K}$. We show that the triple is correctly handled by CHC (ii). Specifically, we assume the inductive hypothesis for all the children of $t$ in $\mathcal{K}$, i.e., for every $j \in [k+m]$, $\mathbb{I}(\mathbf{Pre})(\tau_j, q_j, e_j)$ is true, where the three components comply with items (1)-(3) of Theorem 7.1 applied to the $j^{\mathrm{th}}$ child of $t$ in $\mathcal{K}$. The constraints in the second line of the body of CHC (ii) hold due to Lemma 5.1. The remaining parts of the constraint are also satisfied as detailed in the description of CHC (ii). Hence, with all constraints in the body of the CHC being true, $\mathbb{I}(\mathbf{Pre})(\sigma, q, e)$ must also be true, concluding the proof.                                               □

*Main result of the section.* We conclude the section by linking $C_{\mathrm{pre}}(\mathcal{I}, \mathscr{A}_{\mathrm{pre}})$ to the exit-status problem with precondition by combining Theorem 7.1 with the properties of CHC (iii).

THEOREM 7.2. *Let $(\mathcal{I}, \mathscr{A}_{\mathrm{pre}})$ be an instance of the exit-status problem with precondition. Then, $(\mathcal{I}, \mathscr{A}_{\mathrm{pre}})$ admits a positive answer if and only if the CHC system $C_{\mathrm{pre}}(\mathcal{I}, \mathscr{A}_{\mathrm{pre}})$ is unsatisfiable.*

Proof. Let $\mathcal{I} = (P, k, m, n, Ex)$. We prove the theorem by establishing the equivalent statement: $(\mathcal{I}, \mathscr{A}_{\mathrm{pre}})$ admits a negative answer if and only if the CHC system $C_{\mathrm{pre}}(\mathcal{I}, \mathscr{A}_{\mathrm{pre}})$ is satisfiable.

For the "only if" direction, if $(\mathcal{I}, \mathscr{A}_{\mathrm{pre}})$ admits a negative answer, by definition of exit-status problem with precondition, no $(m, n)$-knitted-tree $\mathcal{K}$ of $P$ with $in(\mathcal{K}) \in L(\mathscr{A}_{\mathrm{pre}})$ has an exit status in *Ex*. Consider an interpretation for $\mathbf{Pre}$ that contains all and only the triples $(\sigma, q, e)$ complying with items (1)-(3) of Theorem 7.1, which correspond to all the triples $(\mathcal{K}, t, \rho)$ described by the items (a)-(c) of the same theorem. By Theorem 7.1, this interpretation satisfies all clauses of $C_{\mathrm{pre}}(\mathcal{I}, \mathscr{A}_{\mathrm{pre}})$ but its query. Also, it satisfies the query because the triple $(\sigma, q, e)$, where $\sigma$ is the log of the root of any $(m, n)$-knitted-tree, cannot satisfy both $e$ and $\psi^F(q)$, showing that $C_{\mathrm{pre}}(\mathcal{I}, \mathscr{A}_{\mathrm{pre}})$ is satisfiable.

For the "if" direction, assume that $C_{\mathrm{pre}}(\mathcal{I}, \mathscr{A}_{\mathrm{pre}})$ is satisfiable. Assume by contradiction that there exists an $(m, n)$-knitted-tree $\mathcal{K}$ of $P$ with $in(\mathcal{K}) \in L(\mathscr{A}_{\mathrm{pre}})$ and $\mathcal{K}$ has an exit status in *Ex*. By applying the "if" direction of Theorem 7.1, we obtain that any interpretation of $\mathbf{Pre}$ that satisfies $C_{\mathrm{pre}}(\mathcal{I}, \mathscr{A}_{\mathrm{pre}})$ except its query contains a triple $(\sigma, q, e)$ where $\sigma$ is the log of $\mathcal{K}$'s root, $q$ is the state of an accepting run of $\mathscr{A}_{\mathrm{pre}}$ on $in(\mathcal{K})$, and $e = true$. This triple clearly violates the query of $\mathbf{Pre}$, hence $C_{\mathrm{pre}}(\mathcal{I}, \mathscr{A}_{\mathrm{pre}})$ cannot be satisfiable.                                               □

# 8 Verifying memory safety

This section presents a sound algorithm for the memory safety problem (defined in §3). We begin by establishing two key theorems that bridge the exit status problem to the memory safety problem, laying the foundation for proving our method's correctness.

The following theorem is an immediate consequence of Lemma 5.4.

**Theorem 8.1.** *If the answer to the exit status problem $(P, k, m, n, \{E\})$ is positive, then the answer to the memory safety problem without precondition for $P$ is negative. The implication still holds if the same precondition is applied to both problems.*

**Theorem 8.2.** *If the answer to the exit status problem $(P, k, m, n, \{O, M, E\})$ is negative, then the answer to the memory safety problem without precondition for $P$ is positive. The implication still holds if the same precondition is applied to both problems.*

Proof. Since no knitted-tree has exit status $O$, by Lemma 5.5, no execution is infinite. Moreover, since no knitted-tree has an exit status in $\{O, M, E\}$, by Lemma 5.4, no execution ends in an error configuration. It follows that all executions end in a final configuration. □

We present our algorithm to solve the memory safety problem without precondition, outlined in Fig. 7. The input is a program $P$, and a natural number $k$ representing the arity of input data trees. The process incorporates two parameters, $m$ and $n$, initialized to default values $m_0$ and $n_0$, which are dynamically incremented during the analysis. The verification process starts by solving the instance $(P, k, m, n, \{E\})$ of the exit status problem, denoted by ExitStatus$(P, k, m, n, \{E\})$, to check for null-pointer dereference errors.

If solving this instance yields a positive answer, according to Theorem 8.1, we conclude a negative answer for the memory safety problem for $P$ (indicating a memory safety violation). If the ExitStatus$(P, k, m, n, \{E\})$ does not yield a positive answer, it does not guarantee memory safety, because the actual values of $m$ and $n$ might not capture all program executions, potentially missing memory safety violations. To address this limitation, we proceed as follows:

(1) We solve the instance ExitStatus$(P, k, m, n, \{M\})$ to check for out-of-memory failure arising from an unsuccessful memory allocation using the **new** statement.
(2) We solve the instance ExitStatus$(P, k, m, n, \{O\})$ to check for label overflow errors.

If both instances yield negative answers, based on Theorem 8.2, we can confirm that $P$ is memory-safe. Otherwise, we increment the corresponding parameter value (either $m$ or $n$) to explore a broader range of program executions. The analysis then restarts with the updated parameter value.

This iterative process may continue indefinitely for two reasons: either the semi-algorithm for solving the exit status problem never returns (due to undecidability), or no value for (at least) one parameter allows us to conclude the analysis when $P$ is memory-safe.

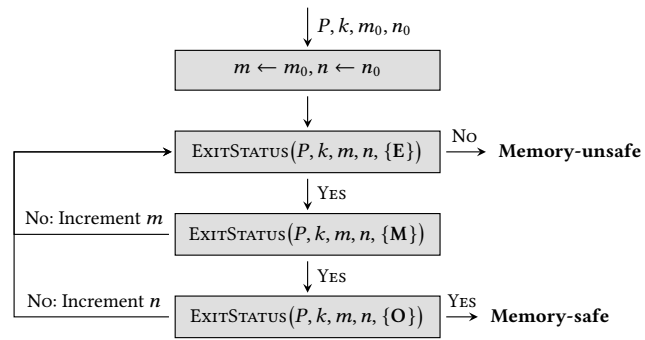

Fig. 7. A sound algorithm for the memory safety problem, where $P$ is a program that takes a $k$-ary data tree as input, and $m_0, n_0 \in \mathbb{N}$ initialize parameters $m$ and $n$.

**Theorem 8.3.** *The procedure in Fig. 7 is a sound solution to the memory safety problem without precondition. I.e., if the procedure terminates, it provides the correct answer to the memory safety problem without precondition.*

*Coping with preconditions.* We now extend the earlier verification process to handle program preconditions. The core structure remains unchanged; we replace each instance of the exit status problem (Fig. 7) with a version that includes a precondition. This involves using the same parameters along with an Sdta that specifies the program's precondition. The correctness of this enhanced verification process is ensured by Theorems 8.1 and 8.2, which already handle preconditions.

## 9   Implementation and Experiments

We have realized a proof-of-concept implementation of our approach, using the CHC solver Z3 [18] and its Python API to build the system of CHCs described in Section 6. The experiments described in this section were performed on an AMD Ryzen 5900X PC running Windows, equipped with 32GB of RAM. As we noticed significant performance differences between different versions of Z3, we report the lowest running times among versions 4.12.0 and 4.12.6. Fig. 8 describes the parameters of the case studies and the size of the corresponding CHC system, when saved in the SMT-LIB format. Table 2 reports the experimental outcomes, discussed in this section.[2]

*Description of the input programs.* We conducted experiments on two heap-manipulating functions, each with a correct version and a memory-unsafe version. The first function, called "List removal" and listed in Appendix A, searches for a given value in a singly-linked list and removes the first node with that value, if any. Singly-linked lists are encoded with $k = 1$, and since the function does not allocate new nodes, we set $m = 0$. The memory-unsafe version skips the initialization of an auxiliary pointer (Line 1 in the source code). Since uninitialized pointers are considered **nil**, this bug leads to a memory violation if the desired value is found in the first node of the list.

The second function is the Bst insertion procedure in Fig. 1b. As it manipulates binary trees and it allocates a new node, we set parameters $k = 2$ and $m = 1$. Its memory-unsafe version consists in moving Line 3 ($p := p'$) at the end of the while-loop. In that way, at the end of the loop both $p$ and $p'$ are **nil**, and a memory violation follows if Lines 12 or 13 are executed.

*Experimental outcomes.* Our tool can thoroughly analyze the four cases under examination, typically within a few seconds, except for two outliers that require 45 and 51 minutes. The parameter $n$ (i.e., number of frames per log) is a critical performance factor, as expected from the encoding. Indeed, some parts of the encoding, such as the predicates responsible for the rewinding process, generate formulas that are quadratic or cubic in $n$.

|  |  |  | Size (MB) | |
|---|---|---|---|---|
| Program | k | m | n=3 | n=4 |
| List removal | 1 | 0 | 0.7 | 2.0 |
| List removal w. bug | 1 | 0 | 0.7 | – |
| Bst insertion | 2 | 1 | 2.0 | 6.0 |
| Bst insertion w. bug | 2 | 1 | 1.6 | – |

Fig. 8.   Parameters and size of the CHC system $C_{ex}(I)$ in SMT-LIB format.

We run each experiment with increasing values of $n$ until either the program is proven correct, or a memory violation is found. For the correct list removal function, $n = 4$ captures all executions. Proving the existence of a terminating execution (unsat of exit status $C$) takes 2 seconds, whereas the absence of memory faults (sat of exit status $E$) takes 31 seconds, and proving the completeness of the analysis (sat of exit status $O$) takes 48 seconds. In contrast, the memory fault in the buggy version of list removal is detected with $n = 3$ in 2 seconds (unsat of exit status $E$). Incidentally, the error in the function simplifies the range of possible executions, making $n = 3$ sufficient to fully explore all possible executions, as confirmed by the "sat" outcome for the exit status $O$.

The Bst insertion function also requires $n = 4$ to be proved correct. Here, we find two performance outliers: 45 minutes to prove the analysis complete and 51 minutes for the guaranteed success of **new** (sat for exit status $M$). This can be partially explained by the fact that the system of CHCs is three times as large as the one for the correct list removal (see Fig. 8). The error in the modified version, however, can be detected with $n = 3$ in 26 seconds.

---

[2]Our encoding uses datatypes, that are not readily supported by other CHC solvers.

| Program | Exit status | Result | | Program | Exit status | Result | |
|---|---|---|---|---|---|---|---|
| | | n=3 | n=4 | | | n=3 | n=4 |
| List removal | O | unsat (00:05) | sat (00:48) | Bst insertion | O | unsat (00:26) | sat (44:53) |
| | C | unsat (00:01) | unsat (00:02) | | C | unsat (00:12) | unsat (01:00) |
| | E | sat (00:06) | sat (00:31) | | E | sat (02:48) | sat (00:17) |
| | M | sat (00:01) | sat (00:01) | | M | sat (00:06) | sat (50:58) |
| List removal w. bug | O | sat (00:07) | – | Bst insertion w. bug | O | sat (03:00) | – |
| | C | unsat (00:01) | – | | C | unsat (00:07) | – |
| | E | **unsat (00:02)** | – | | E | **unsat (00:26)** | – |
| | M | sat (00:01) | – | | M | sat (00:04) | – |

Table 2. Experimental results on memory safety. "sat" and "unsat" refer to the CHC system. Hence, "sat" means that the corresponding exit status is *not* achieved by any program execution. Time is in mm:ss format and does not include the time to generate the system of CHCs.

The encoding and tool implementation are currently unoptimized. Potential optimizations are discussed in §12 (future work).

## 10 Verifying Other Properties

This section presents in ascending order of complexity other program properties, besides memory safety, that can be addressed using the framework presented in this paper. Due to space constraints, we provide a high-level description here, with formal details to be covered in a future extension.

*Checking simple contracts.* When a tree data structure represents a set of elements, standard operations include *insertion*, *deletion*, and *search*. It is crucial that these operations maintain the tree's expected properties. Specifically, nodes not directly involved in an operation should remain unaffected in terms of their node allocation status and the data they store. For insertion or deletion, the operation should add or remove the specified node, respectively. We refer to these requirements as the **insertion contract** and the **deletion contract**. Search operations should not modify the data in any way, a property we call the **read-only contract**. We can address the insertion, deletion, and read-only contracts using Sdta-based approaches. The idea is to design an Sdta that works on knitted-trees and accepts them if they *violate* the property of interest. Consider the read-only contract. Its Sdta operates bottom-up on the input knitted-tree, checking each node to ensure that its log does not contain unavailable frames where the field *active* changes state, or the pointer fields or data fields are updated. If either of these conditions is violated, the Sdta stores this information in its state. Upon reaching the root of the knitted-tree, the Sdta accepts it if a violating node was found. We can then intersect this Sdta with the one expressing the precondition, and simulate them together on the set of knitted-trees, following the CHC construction in §7. A similar approach can be designed for the other contracts, ensuring proper insertion, deletion, and search behaviors.

*Checking the output tree.* When dealing with tree data structures, ensuring that the program's execution consistently produces a valid tree is of utmost importance. Firstly, we aim to guarantee that the heap's underlying structure resulting from any execution is indeed a tree, referred to as the **tree-structure postcondition**. To check this condition on a single knitted-tree, one could start from the node of the output tree identified by $\widehat{p}$ and recursively mark all descendants that are active at the end of the execution as reachable. Note, however, that the children of a node in the output tree may be different from its *structural* children (i.e., different from the children in the knitted-tree). If a node's log contains the symbolic instruction ⟨*pfield* := *p*⟩, it indicates that a link of the current node was reassigned to the current value of the pointer *p*. The postcondition check must then trace back the value of *p*, similar to the rewind operation described in §4.3. If, at any time, the postcondition verification finds that a node is reachable through two different paths, it will declare that the output is not a tree.

The procedure sketched above can be simulated by a non-deterministic Sdta $\mathcal{A}_{\neg tree}$ that accepts all knitted-trees whose output is *not* a tree. Intuitively, $\mathcal{A}_{\neg tree}$ would guess which nodes are reachable and which pointers are traced along the lace, and then, in a single bottom-up pass, check the correctness of the guesses. At that point, the CHCs in Figure 6 can be easily adapted to run $\mathcal{A}_{\neg tree}$ on *all* knitted-trees of a given program, and so verify that all executions produce a tree.

A similar approach can be used to check other postconditions, such as the fact that the output is a specific kind of tree (e.g., a Bst, a red-and-black tree, etc.). In that case, one would start by encoding such property as an Sdta $\mathcal{A}$ that checks the negated property on data trees. To run this Sdta on a single knitted-tree, one would again start from the node that is the root of the output tree and label it with one of the accepting states of $\mathcal{A}$. Then, one would recursively visit all descendants in the output tree, identified by the same pointer-tracing procedure mentioned above, and label them with the appropriate states of $\mathcal{A}$ according to its transition relation. If this process manages to build a fully coherent accepting run of $\mathcal{A}$ on the output tree, it has been proved that it violates the original property. As described above for $\mathcal{A}_{\neg tree}$, this algorithm can be emulated by a non-deterministic Sdta on knitted-trees, and then turned into a system of CHCs analogous to Figure 6.

*Deductive Verification.* Our methodology is specifically tailored to establish the correctness of procedures with a time complexity linearly proportional to the number of nodes in input lists or trees. This limitation arises because we can only handle a bounded number of passages through each node. While this constraint suits many common procedures in the literature that operate in linear time, our approach can also be used in deductive verification. To apply our approach in that context, one decomposes the verification process into more manageable conditions, describing pre- and postconditions of specific segments of code that, in the worst case, can be executed in linear time. Successfully proving these verification conditions serves as a collective affirmation, establishing the overall correctness of the entire procedure under analysis.

## 11   Related Work

Our work is related to many works in the literature in different ways. Here we focus on those that seem to be closest to the results presented in this paper.

Our approach uses CHC engines for backend analysis, aligning with other program verification methods [13, 22, 24, 25, 27, 29, 30, 33, 35, 39, 40, 52]. CHCs serve as an intermediate verification language, allowing focus on proof rules while solvers implement algorithms within a standard framework. A major challenge is encoding heap-allocated mutable data structures. While array theory is often used (e.g., [16, 41]), it can result in complex CHCs. Our approach uses simple theories for basic data types, avoiding array theory unless necessary. Traditional heap program analysis often uses abstractions like shape analysis [64] to scale. Refinement types and invariants can be used to transform complex data structures, avoiding array theory (e.g. [11, 38, 53, 63]). This method often leads to over-approximation in CHCs, potentially causing false positives, as it replaces heap operations with local object assertions. While this may not capture global invariants, it can enable efficient verification for programs where local invariants suffice. A recent proposal suggests using an SMT-LIB theory of heaps for CHCs to standardize heap data-structure representation [21].

Our technique intersects with tree automata, automata with auxiliary storage and bounded tree-width graphs representing their executions. It also relates to Courcelle's theorem proof, which reduces analysis to tree automata emptiness [26]. Alur and Madhusudan [3] represented pushdown automata executions with nested words, later extended for multistack and distributed automata [46]. We apply similar concepts to tree-manipulating programs, using laces as a graph representation of program executions, where nodes are frames and edges are *next* and *prev* frame fields. La Torre et al. [43] proposed a tree encoding for multistack pushdown automata executions, comparable to our lace encoding. In both cases, these graph representations of executions constitute a tree decomposition, witnessing their bounded treewidth. While [43] uses tree automata emptiness to analyze all executions of multistack pushdown automata and solve their reachability problem, we leverage CHC solvers. This approach allows us to employ a tree automata-like method with the added advantage of reasoning about frame data. Similarly to [31], we use automata for program analysis. However, whereas [31] employs counterexample-guided abstraction refinement to iteratively refine an overapproximation of a program's error paths that is represented by an automaton, we use precise a knitted-tree representation and leverage CHC solvers for approximation and refinement.

A line of work in program verification focuses on developing decidable approaches for bounded-pass heap-manipulating programs. While our work also considers bounded-pass programs, it offers a broader range of verifiable properties compared to these approaches, potentially at the cost of decidability. Mathur et al. [51] demonstrate memory safety decidability under specific conditions: forest-like initial heaps and single-pass data structure traversal. Their results build on earlier findings on uninterpreted coherent programs [49, 50]. While that approach handles memory freeing, the support for more complex postconditions is left for future work. Alur and Černý [2] reduce assertion checking of single-pass list-processing programs to questions on data string transducers. Its decidability stems from using a single advancing variable, limiting it compared to Mathur et al.'s work, which supports multiple variables for pointer updates. This approach does not directly address memory safety or heap shape changes, and is limited to reasoning about data ordering and equality. It also does not handle explicit memory freeing.

Heap verification has been extensively studied, with research spanning decidable logics including first-order logics with reachability [45], Lisbq in the Havoc tool [44], and separation logic fragments [8, 20, 59]. Some approaches interpret bounded treewidth data structures on trees [34, 47]. While these logics are often restricted, others with undecidable validity employ effective strategies using heuristics, lemma synthesis, and programmer annotations [5–7, 9, 10, 12, 14, 15, 19, 36, 54, 56, 58, 60, 65]. In contrast, our knitted-tree encoding promotes a separation of concerns, offloading the algorithmic burden to the underlying CHC solver.

## 12 Conclusions and Future Works

Verifying programs that manipulate tree data structures often relies on intricate, manual proofs, hindering automation and generalization. This paper introduces an automated technique to overcome these challenges. Our core contribution lies in a unified approach that leverages automata and logics to handle various tree data structures. The foundation of our method is the knitted-tree encoding, which ingeniously combines program inputs, outputs, and intermediate states into a single structure. This encoding, coupled with its compositional properties, allows us to effectively represent knitted trees using CHCs. By translating verification into the well-studied domain of CHC satisfiability, we benefit from efficient off-the-shelf CHC solvers. We demonstrated our technique's applicability to the memory safety problem and discussed its potential for broader verification tasks. Our technique shows promise for automating verification of linear-time procedures on tree-like data structures, and can also assist in deductive verification of higher-complexity programs.

*Future work.* Our primary focus will be on enhancing the practical effectiveness of our approach through performance optimization. A key area for improvement is the size of the logs used in our current implementation of knitted-trees. Our current approach follows the theoretical description and does not optimize this parameter. We envision using summaries (see for example [28, 42]) to compress the internal steps within the lace. This optimization could potentially reduce the log size to just two frames for several procedures involving common tree data structures, which typically navigate the input tree along a descending path. As demonstrated in our experimental results, analyses carried out with short logs exhibit significant performance improvements.

Regarding the applicability to data structures beyond trees, we mention in the related work section how the graphs underlying each lace associated with a knitted-tree exhibit bounded treewidth, with the decomposition derived from the knitted-tree itself. This suggests that our technique may be well-suited to a broader range of data structures, including doubly-linked lists and trees with parent pointers, and generally, structures with bounded treewidth and canonical tree decompositions.

While program logics for heap verification, such as separation logics [57, 62] and related first-order logics [55], are well-established, we use Symbolic Data Tree Automata as a lower-level mechanism. Future work could explore converting properties from these higher-level logics to equivalent SDTAs, combining the intuitive nature of established logics with the advantages of our automata-based method.

Another promising application of our approach is in developing a syntax-guided synthesis (SyGuS) method [1] for procedures that manipulate tree data structures, with preconditions and postconditions expressed by SDTAs. This method would enable the generation of code that is correct by construction using CHCs, addressing similar challenges to those discussed in [61].

## A List Removal Example

This appendix reports the source code of one of the case studies from Section 9. This function removes the first node whose data field contains a value equal to *key*, assuming that the pointer variable *head* initially points at the head of a singly linked list.

```
      pointer head, prev
      int key, value
0 :  prev := head ;
1 :  while (head ≠ nil) do
2 :      value := head→data ;
3 :      if (key = value) then
4 :          head := head→next ;
5 :          prev→next := head ;
6 :          exit ;
          else
7 :          prev := head ;
8 :          head := head→next ;
          fi ;
      od ;
9 :  exit ;
```

## B Additional Proofs

LEMMA 5.1. *For all labels* $\sigma, \tau \in L(\mathcal{S}_{\mathcal{K}})$ *and indices* $j \in [k+m]$, *if there exists a knitted-tree prefix where* $\sigma$ *and* $\tau$ *are the logs of a node and its* $j$*-th child respectively, then* consistent_child$(\tau, j, \sigma)$ *holds.*

PROOF. We follow the definition of *consistent_child* presented in Appendix C.4. The *consistent_first_frames*$(\tau, j, \sigma)$ predicate holds true because it follows the definition of the *active* and *active_child* fields in the first frame of every node, as detailed in Sec. 4.3.1. The next two blocks in the definition of *consistent_child* are responsible for checking that every pair of frames connected by the *prev* and *next* fields and belonging to different labels encodes a step in the current instruction, according to the rules detailed in §4.3. The two connected frames may represent a step downward (first block) or upward (second block) in the knitted-tree.

Consider the downward direction, as the other case is symmetrical. A downward connection between the frames $\sigma^a$ and $\tau^b$ is detected in two cases:

(1) First, if $\sigma^a$ is occupied (i.e., not available), linked to $\tau^b$ through the *next* field, and $\tau^b$ is also occupied. Then, we apply the predicate $\Psi_{Down}$ to check that the step is valid. Note that it is possible that in a valid knitted-tree $\tau^b$ is available, even though $\sigma^a$.*next* points to it, but only when $\sigma^a$ is the last frame of the lace.

(2) Second, when $\tau^b$ is occupied and linked to $\sigma^a$ via its *prev* field. In that case, $\sigma^a$ *must* be occupied in any valid knitted-tree, and we again check the correctness of the step using $\Psi_{Down}$.

□

## C   Detailed Encoding

In this section, we describe in detail the predicates and functions appearing as constraints in the CHCs that recognize knitted-trees. Consider again the system of CHCs described in Figure 5. Those CHCs are based on the predicates *first_frame*, *start*, *consistent_child*, $\Psi_{Internal}$, $\Psi_{Down}$, $\Psi_{Up}$, and *label_exit*, each described in one of the following sections.

### C.1   The First Frame of Non-Root Nodes

The predicate in this section is meant to constrain the first frame of every non-root node of a knitted-tree. The only property to be enforced is a relationship between the fields *active* and *active_child*: an auxiliary node (*active = false*) that is not the root must be a leaf of the knitted-tree. Hence, no allocation of new nodes can be performed in this node. This is enforced by setting the *active_child_j* flags to *true* for all indices $j \in [k+1, k+m]$. Vice versa, all active nodes start with $m$ auxiliary children with indices $k+1, \ldots, k+m$, available for allocation. We obtain the following predicate:

$$\underline{first\_frame}(f) \stackrel{\text{def}}{=} \left( f.active \wedge \bigwedge_{j \in [k+1,k+m]} \neg f.active\_child_j \right) \vee$$
$$\left( \neg f.active \wedge \bigwedge_{j \in [k+1,k+m]} f.active\_child_j \right).$$

### C.2   The Start of the Lace

The predicate *start* sets constraints on the first two frames at the root of any knitted-tree. These frames must agree on the fields *active* and *key*. The *prev* field of the second frame points to itself through the value $(-, 2)$. For non-empty input trees, the first frame must be active and the second frame makes $\widehat{p}$ point to the root node, and *active_child* is non-deterministic on the first $k$ children, constraining only the children from index $k+1$ to $k+m$ as inactive. For empty input trees, the first frame is inactive and the second frame sets $\widehat{p}$ to **nil**, and *active_child* requires all children to be inactive. In both cases, $pc^2$ is set to the first program statement label, with all the other pointer variables set to **nil**.

$$
\begin{aligned}
\underline{start}(\sigma) \stackrel{\text{def}}{=} \; &initial(\sigma) \wedge \\
&\sigma.active^2 = \sigma.active^1 \; \wedge \; \sigma.key^2 = \sigma.key^1 \wedge \\
&\sigma.pc^2 = 0 \; \wedge \; \bigwedge_{p \in PV_P \setminus \{\widehat{p}\}} \sigma.isnil_p^2 \wedge \\
&\Big[ \left( \sigma.active^1 \; \wedge \; \sigma.instr^2 = \langle \widehat{p} := \mathbf{here} \rangle \; \wedge \; \neg \sigma.isnil_{\widehat{p}}^2 \wedge \right. && \text{Non-empty input tree} \\
&\quad \left. \bigwedge_{j \in [k+1,k+m]} \neg \sigma.active\_child_j^2 \right) \; \vee \\
&\quad \left( \neg \sigma.active^1 \wedge \sigma.instr^2 = \textsc{nop} \; \wedge \; \sigma.isnil_{\widehat{p}}^2 \wedge \right. && \text{Empty input tree} \\
&\quad \left. \bigwedge_{j \in [k+m]} \neg \sigma.active\_child_j^2 \right) \Big],
\end{aligned}
$$

where the *initial* predicate describes the distinguishing feature of the initial frame of a lace – namely, having itself as predecessor:

$$\underline{initial}(\sigma) \stackrel{\text{def}}{=} \neg \sigma.avail^2 \wedge (\sigma.prev^2 = (-, 2)).$$

## C.3 Lace Termination

The function $label\_exit(\sigma)$ returns the exit status of any frame in $\sigma$, assuming that at most one frame in a label is terminal. In the following, we denote by $P(i)$ the instruction located at program counter $i$ in the program $P$.

$$\underline{continues}(f) \stackrel{\text{def}}{=} \neg f.avail \wedge frame\_exit(f) = N$$

$$\underline{frame\_exit}(f) \stackrel{\text{def}}{=} \begin{cases} C & \text{if } P(f.pc) = \textbf{exit} \\ E & \text{if } f.instr = \textsc{err} \\ M & \text{if } f.instr = \textsc{oom} \\ N & \text{otherwise.} \end{cases}$$

$$\underline{label\_exit}(\sigma) \stackrel{\text{def}}{=} \begin{cases} C & \text{if } \bigvee_{i \in [2,n]} \left( \neg\sigma.avail^i \wedge frame\_exit(\sigma^i) = C \right) \\ E & \text{if } \bigvee_{i \in [2,n]} \left( \neg\sigma.avail^i \wedge frame\_exit(\sigma^i) = E \right) \\ M & \text{if } \bigvee_{i \in [2,n]} \left( \neg\sigma.avail^i \wedge frame\_exit(\sigma^i) = M \right) \\ O & \text{if } \neg\sigma.avail^{n+1} \\ N & \text{otherwise.} \end{cases}$$

$$\underline{label\_exit}(\sigma, Ex) \stackrel{\text{def}}{=} \left( label\_exit(\sigma) \in Ex \right)$$

## C.4 Checking Parent-Child Consistency

The $consistent\_child(\tau, j, \sigma)$ predicate checks that the label $\tau$ of the $j$-th child is consistent with the label $\sigma$ of the parent. It uses the predicates $consistent\_first\_frames$, $\Psi_{Down}$, and $\Psi_{Up}$, defined later.

$$\underline{consistent\_child}(\tau, j, \sigma) \stackrel{\text{def}}{=} consistent\_first\_frames(\tau, j, \sigma) \wedge$$

Consistency of steps, when going down:

$$\bigwedge_{\substack{a \in [2,n] \\ b \in [2,n+1]}} \left[ \left( \left( \neg\sigma.avail^a \wedge \sigma.next^a = (j, b) \wedge \neg\tau.avail^b \right) \vee \right.\right.$$
$$\left.\left. \left( \neg\tau.avail^b \wedge \tau.prev^b = (\uparrow, a) \right) \right) \rightarrow \Psi_{Down}(\sigma^{\leq a}, j, \tau^{<b}, \tau^b) \right] \wedge$$

Consistency of steps, when going up:

$$\bigwedge_{\substack{a \in [2,n] \\ b \in [2,n+1]}} \left[ \left( \left( \neg\tau.avail^a \wedge \tau.next^a = (\uparrow, b) \wedge \neg\sigma.avail^b \right) \vee \right.\right.$$
$$\left.\left. \left( \neg\sigma.avail^b \wedge \sigma.prev^b = (j, a) \right) \right) \rightarrow \Psi_{Up}(\tau^{\leq a}, j, \sigma^{<b}, \sigma^b) \right].$$

It is worth pointing out that different child labels may be consistent with the same parent label, just like two different parent labels may be consistent with the same child label. First, $consistent\_child$ only constrain frames that are directly involved in an interaction between parent and child, i.e., pairs of frames belonging to these different nodes and adjacent in the lace. Now, consider a pair of frames that are adjacent in the lace and belong to two different nodes (parent and child). Here is a description of how $consistent\_child$ constrains each field in those frames:

*avail*: this field must be false, because frames involved in an interaction belong to the lace;
*active*: this field is directly constrained only in the first frame of each label, where its value must follow the backbone rules described in Section 4.1; this is ensured by the auxiliary predicate

*consistent_first_frames*; the value in the subsequent frames of each label is only constrained indirectly, by the semantics of the **new** and **free** instructions;

*key*: the default constraint for this field is to have the same value as the preceding frame in the same log; the only exception is the last step of the instruction $p \rightarrow data := exp$, when the current value of $p$ has been ascertained via rewinding: in such a step the value of the *key* field of the new frame takes the value of the expression *exp*;

*pc*: the program counter must be consistent with the corresponding *pc* on the other side of the interaction; i.e., the two pc's must be equal if a rewind operation is ongoing, and otherwise the instruction being executed is complete and the pc advances according to the control-flow graph of $P$;

*d*: the value of this field in the two frames of an interaction is the same, except in the last step of an instruction of the type $d := p \rightarrow \mathbf{data}$, when the current value of $p$ has been ascertained via rewinding;

$upd_p$: the auxiliary predicates $\Psi_*$ constrain these fields to follow the rules described on page 9;

$isnil_p$: this field generally keeps its value in an interaction, except in the last step of an assignment to $p$; specifically, it can switch from *true* to *false* in the last step of $p := q$ or $p := q \rightarrow pfield$, when $q$ is not **nil** and rewinding was needed to find its current value; moreover, it can switch from *false* to *true* in the last step of $p := q \rightarrow pfield$, when $q$ is not **nil**, rewinding was needed, and then $q \rightarrow pfield$ is found to be **nil**;

*instr*: the symbolic instruction in the second frame of an interaction is dictated by the program instruction being executed and by the meaning of each symbolic instruction; for example, during a rewind process, the symbolic instructions $\text{RWD}_i$ and $\text{RWD}_{i,p}$ are used; the end of a rewinding may be marked by a symbolic instruction of the types $\langle pfield := p \rangle$, $\langle pfield := \mathbf{nil} \rangle$, or $\langle p := \mathbf{here} \rangle$; in some cases, the final symbolic instruction is simply NOP;

*next* and *prev*: in a pair of interacting frames these fields must connect the two frames with each other; i.e., the *next* field of the first frame points to the second frame, and the *prev* field of the second frame points to the first one.

The following auxiliary predicates are used within *consistent_child*. The *consistent_first_frames* predicate describes the initial value of the *active* field, that starts as true on all nodes of the knitted-tree that correspond to nodes of the input tree, and false elsewhere.

$$\underline{consistent\_first\_frames}(\tau, j, \sigma) \stackrel{\text{def}}{=} \big(initial(\sigma) \vee \sigma.active^1\big) \,\wedge\, \big(\tau.active^1 \rightarrow \sigma.active^1\big) \,\wedge$$
$$\big(j > k \rightarrow \neg\tau.active^1\big) \,\wedge\, \sigma.active\_child_j^1 = \tau.active^1 \,.$$

We now present the three predicates that describe the next frame in the lace, depending on the direction of the next step. The first predicate, $\Psi_{Internal}$, describes internal steps. It invokes the general predicate *step* that checks a single step in a lace, passing two adjacent frames from the same log. In addition, it makes sure that the $upd_p$ flags are all false after an internal step, according to the meaning of those flags, as described on page 9.

In the following formulas, for a label $\sigma$ and a frame $f$ we write $\sigma \cdot f$ to denote the label obtained by the replacing the first available frame of $\sigma$ with $f$.

$$\underline{\Psi_{Internal}}(\sigma, f) \stackrel{\text{def}}{=} len(\sigma, a) \wedge continues(\sigma^a) \wedge \sigma.next^a = (-, a+1) \,\wedge$$
$$f.prev = (-, a) \wedge step(\sigma, a; \sigma \cdot f, a+1) \wedge \bigwedge_{p \in PV_P} \neg f.upd_p \,.$$

The following predicates $\Psi_{Down}$ and $\Psi_{Up}$ describe the next frame in the lace, when the current step involves a change of node, either from a node down to its $j$-th child or from a node up to its parent. In particular, these predicates constrain the *upd* flags in the new frame to respect their meaning, as described on page 9.

Consider how the last two lines in $\Psi_{Up}(\tau, j, \sigma, f)$ relate to the intended behavior of the *upd* flags. In that context, $\tau^a$ is a frame in the $j$-th child of a parent node with log $\sigma$, and the next frame in the lace is $f$, to be appended on top of $\sigma$. Note that in this situation the previous frame in the log of the parent (i.e., $\sigma^{b-1}$) is followed in the lace by a frame in the $j$-th child. Now, for every pointer variable $p \in PV_P$, the $upd_p$ flag when going back to the parent is true iff $p$ is not **nil** at that time and we observe a marker for an update to $p$ within the frames of the child located between the step down and the step back up. If the update to $p$ was performed in one of the frames of the child, the marker is a direct symbolic instruction of the type $\langle p := \textbf{here} \rangle$. If instead the update occurs further below in the subtree rooted in the child, the marker observed in the child is a $upd_p$ field set to true, as encoded in the last lines of $\Psi_{Up}$.

$$\underline{\Psi_{Down}}(\sigma, j, \tau, f) \stackrel{\text{def}}{=} len(\sigma, a) \wedge continues(\sigma^a) \wedge \sigma.next^a = (j, b) \wedge len(\tau, b - 1) \wedge$$
$$f.prev = (\uparrow, a) \wedge step(\sigma, a; \tau \cdot f, b) \wedge$$
$$\bigwedge_{p \in PV_P} \Big[ f.upd_p \leftrightarrow \Big( \neg f.isnil_p \wedge b > 2 \wedge \tau.next^{b-1} = (\uparrow, a') \wedge$$
$$\Big( \bigvee_{c \in [a', a]} \big( \sigma.instr^c = \langle p := \textbf{here} \rangle \big) \vee \bigvee_{c \in [a'+1, a]} \sigma.upd_p^c \Big) \Big) \Big]$$

$$\underline{\Psi_{Up}}(\tau, j, \sigma, f) \stackrel{\text{def}}{=} len(\tau, a) \wedge continues(\tau^a) \wedge \tau.next^a = (\uparrow, b) \wedge len(\sigma, b - 1) \wedge$$
$$f.prev = (j, a) \wedge step(\tau, a; \sigma \cdot f, b) \wedge$$
$$\bigwedge_{p \in PV_P} \Big[ f.upd_p \leftrightarrow \Big( \neg f.isnil_p \wedge b > 2 \wedge \sigma.next^{b-1} = (j, a') \wedge$$
$$\Big( \bigvee_{c \in [a', a]} \big( \tau.instr^c = \langle p := \textbf{here} \rangle \big) \vee \bigvee_{c \in [a'+1, a]} \tau.upd_p^c \Big) \Big) \Big].$$

## C.5 Individual Statements

The predicate $step(\sigma, a; \tau, b)$ holds if $\tau^b$ is the logically correct frame to follow $\sigma^a$ in a lace. The structure of this predicate comprises an implication for each type of instruction in our programming language, except for **exit**. The **exit** instruction does not produce a step, because just having the program counter point to it identifies a terminating execution.

$$\underline{step}(\sigma, a; \tau, b) \stackrel{\text{def}}{=}$$

$$
\begin{aligned}
P(\sigma.pc^a) = \mathbf{skip} \quad &\rightarrow step\_skip(\sigma^a, \tau^b) \wedge \\
P(\sigma.pc^a) = (p \coloneqq \mathbf{nil}) \quad &\rightarrow step\_assgn\_nil(\sigma^a, \tau^b, p) \wedge \\
P(\sigma.pc^a) = (d \coloneqq exp) \quad &\rightarrow step\_assgn\_exp(\sigma^a, \tau^b, d, exp) \wedge \\
P(\sigma.pc^a) = (d_{\text{bool}} \coloneqq (p = q)) \quad &\rightarrow step\_assgn\_cond(\sigma, a, \tau, b, d_{\text{bool}}, p, q) \wedge \\
P(\sigma.pc^a) = (p \coloneqq q) \quad &\rightarrow step\_assgn\_ptr(\sigma, a, \tau, b, p, q) \wedge \\
P(\sigma.pc^a) = (p \coloneqq q \rightarrow pfield) \quad &\rightarrow step\_assgn\_from\_field(\sigma, a, \tau, b, p, q, pfield) \wedge \\
P(\sigma.pc^a) = (p \rightarrow pfield \coloneqq q) \quad &\rightarrow step\_assgn\_to\_field(\sigma, a, \tau, b, p, q, pfield) \wedge \\
P(\sigma.pc^a) = (p \rightarrow \mathbf{data} \coloneqq exp) \quad &\rightarrow step\_assgn\_to\_data(\sigma, a, \tau, b, p, exp) \wedge \\
P(\sigma.pc^a) = (d \coloneqq p \rightarrow \mathbf{data}) \quad &\rightarrow step\_assgn\_to\_var(\sigma, a, \tau, b, d, p) \wedge \\
P(\sigma.pc^a) = (\mathbf{new}\ p) \quad &\rightarrow step\_new(\sigma^a, \tau^{b-1}, \tau^b, p) \wedge \\
P(\sigma.pc^a) = (\mathbf{free}\ p) \quad &\rightarrow step\_free(\sigma, a, \tau, b, p) \wedge \\
\left(P(\sigma.pc^a) = (\mathbf{while}\ p = q)\ \vee \right. & \\
\left. P(\sigma.pc^a) = (\mathbf{if}\ p = q)\right) \quad &\rightarrow step\_cmp\_ptr(\sigma, a, \tau, b, p, q, \mathit{false}) \wedge \\
\left(P(\sigma.pc^a) = (\mathbf{while}\ \neg (p = q))\ \vee \right. & \\
\left. P(\sigma.pc^a) = (\mathbf{if}\ \neg (p = q))\right) \quad &\rightarrow step\_cmp\_ptr(\sigma, a, \tau, b, p, q, \mathit{true}) \wedge \\
\left(P(\sigma.pc^a) = (\mathbf{while}\ r(exp_1, \ldots, exp_l))\ \vee \right. & \\
\left. P(\sigma.pc^a) = (\mathbf{if}\ r(exp_1, \ldots, exp_l))\right) \quad &\rightarrow step\_local\_branch(\sigma, a, \tau, b, r, exp_1, \ldots, exp_l)\,.
\end{aligned}
$$

*C.5.1 Default Values.* Regardless of any specific instruction, most fields in any new frame of the lace are set to their default values, as described in Section 4.3.2. The following *default* predicate represents the full case, where all fields that have default values are constrained. In the subsequent predicates, we add subscripts to *default* to specify which fields follow the default rules.

$$
\begin{aligned}
\underline{default}(f^{prev}, f^{below}, f) \stackrel{\text{def}}{=} \neg f.avail\ \wedge \quad &\textit{avail} \\
f.active = f^{below}.active\ \wedge \quad &\textit{active} \\
f.key = f^{below}.key\ \wedge \quad &\textit{key} \\
\textstyle\bigwedge_{d \in DV_P} f.d = f^{prev}.d\ \wedge \quad &\textit{d} \\
\textstyle\bigwedge_{p \in PV_P} f.isnil_p = f^{prev}.isnil_p\ \wedge \quad &\textit{isnil} \\
f.instr = \textsc{nop}\ \wedge \quad &\textit{instr} \\
f.pc = f^{prev}.pc\ \wedge \quad &\textit{pc} \\
default_{active\_child}(f^{prev}, f^{below}, f)\,. \quad &\textit{active\_child}
\end{aligned}
$$

The default for the *active_child* flags requires a little more care:

$$\underline{default_{active\_child}}(f^{prev}, f^{below}, f) \stackrel{\text{def}}{=}$$

$$f.prev = (j, *) \rightarrow \Big( f.active\_child_j = f^{prev}.active \; \wedge$$

$$\bigwedge_{j \neq i} \big( f.active\_child_i = f^{below}.active\_child_i \big) \Big) \; \wedge$$

$$f.prev \neq (j, *) \rightarrow \bigwedge_{i \in [k+m]} \big( f.active\_child_i = f^{below}.active\_child_i \big) \,.$$

*C.5.2 Local Instructions.* As our first type of local instruction, we describe the *step_assgn_nil* predicate. As explained earlier, its encoding simply pushes a new frame $f_2$ on the current node, with the *isnil* flag set to *true*. All the other fields of the new frame take their default value, except the program counter, which advances to the next instruction:

$$\underline{step\_assgn\_nil}(f_1, f_2, p) \stackrel{\text{def}}{=} \; (f_1.next = -) \; \wedge \hspace{2cm} p := \text{nil}$$

$$f_2.isnil_p \; \wedge \; \bigwedge_{q \in PV_P \setminus \{p\}} (f_2.isnil_q = f_1.isnil_q) \; \wedge$$

$$advance\_pc(f_1, f_2) \wedge default_{avail, active, key, d, instr, active\_child}(f_1, f_1, f_2).$$

The following auxiliary predicate encodes the advancement of the program counter:

$$\underline{advance\_pc}(f_1, f_2) \stackrel{\text{def}}{=} \; (f_2.pc = succ(f_1.pc)).$$

Next, we present the predicate that encodes the **new** instruction. Note that a failed **new** is only justified if all children of the current node with position in $[k+1, k+m]$ are active (i.e., currently allocated).

$$\underline{step\_new}(f, f', f'', p) \stackrel{\text{def}}{=} \hspace{4cm} \text{new } p$$

Case 1: Out-of-memory error

$$\Big( \bigwedge_{j \in [k+1, k+m]} f.active\_child_j \; \wedge \; f.next = (-, *) \; \wedge \; f''.instr = \text{OOM} \; \wedge$$

$$default_{avail, active, key, d, pc, isnil, active\_child}(f, f', f'') \Big) \quad \vee$$

Case 2: Normal case

$$\Big( \bigvee_{j \in [k+1, k+m]} \big( \neg f.active\_child_j \; \wedge \; \bigwedge_{i \in [k+1, j-1]} f.active\_child_i \; \wedge \; f.next = (j, *) \big) \; \wedge$$

$$\neg f''.isnil_p \; \wedge \; \bigwedge_{q \in PV_P \setminus \{p\}} \big( f''.isnil_q = f.isnil_q \big) \; \wedge$$

$$f''.instr = \langle p := \textbf{here} \rangle \; \wedge \; f''.active \; \wedge \; advance\_pc(f, f'') \; \wedge$$

$$default_{avail, key, d, active\_child}(f, f', f'') \Big).$$

The following predicate handles assignments of arbitrary data expressions to data variables.

$$\underline{step\_assgn\_exp}(f_1, f_2, d, exp) \stackrel{\text{def}}{=} \hspace{4cm} d := exp$$

$$\big( f_1.next = (-, *) \big) \; \wedge \; f_2.d = exp[d' \mapsto f_1.d']_{d' \in DV_P} \; \wedge$$

$$\bigwedge_{d' \in DV_P \setminus \{d\}} \big( f_2.d' = f_1.d' \big) \; \wedge \; advance\_pc(f_1, f_2) \wedge$$

$$default_{avail, active, key, isnil, instr, active\_child}(f_1, f_1, f_2) \,.$$

Finally, the following is the straightforward encoding of the **skip** statement.

$$\underline{step\_skip}(f_1, f_2, p) \stackrel{\text{def}}{=} \big(f_1.next = (-, *)\big) \;\wedge\; advance\_pc(f_1, f_2) \;\wedge\; \qquad \text{skip}$$
$$default_{avail, active, key, d, isnil, instr, active\_child}(f_1, f_1, f_2).$$

*C.5.3 Walking Instructions.* A more complex instruction is the assignment of the form $p := q$, which may require rewinding the lace to find the node currently pointed by $q$. In fact, the encoding distinguishes three cases: (1) when $q$ is **nil**, (3) when $q$ points elsewhere, and we need to start or keep rewinding the lace, (2) when $q$ points to the current node (i.e., the node with label $\sigma$). The predicates governing the rewinding operation are presented later in Section C.7.

$$\underline{step\_assgn\_ptr}(\sigma, a, \tau, b, p, q) \stackrel{\text{def}}{=} \qquad\qquad\qquad p := q$$
$$\qquad \text{Case 1: } q \text{ is } \mathbf{nil}; \text{ use the encoding for } p := \mathbf{nil}$$
$$\Big(\sigma^a.isnil_q \;\wedge\; step\_assgn\_nil(\sigma^a, \tau^b, p)\Big) \quad\vee$$
$$\qquad \text{Case 2: } q \text{ points elsewhere; start or keep rewinding}$$
$$rewind(\sigma, a, \tau, b, q) \quad\vee$$
$$\qquad \text{Case 3: } q \text{ points to the current node}$$
$$\Big(stop\_rewind(\sigma, a, q) \;\wedge\; set\_ptr\_here(\sigma^a, \tau^b, p)\Big).$$

The auxiliary predicate *set_ptr_here* pushes a frame on the current node with $\langle p := \mathbf{here}\rangle$ and updates *isnil* accordingly.

$$\underline{set\_ptr\_here}(f_1, f_2, p) \stackrel{\text{def}}{=} \big(f_1.next = (-, *)\big) \;\wedge\; advance\_pc(f_1, f_2) \;\wedge\;$$
$$f_2.instr = \langle p := \mathbf{here}\rangle \;\wedge\;$$
$$\neg f_2.isnil_p \;\wedge\; \bigwedge_{q \in PV_P \setminus \{p\}}\big(f_2.isnil_q = f_1.isnil_q\big) \;\wedge\;$$
$$default_{avail, active, key, d, active\_child}(f_1, f_2).$$

The following predicates encodes the statements of the form $p \rightarrow pfield := q$.

$$\underline{step\_assgn\_to\_field}(\sigma, a, \tau, b, p, q, pfield) \stackrel{\text{def}}{=} find\_or\_fail(\sigma, a, \tau, b, p) \;\vee \qquad p \rightarrow pfield := q$$
$$\qquad p \text{ points to the current node:}$$
$$\Big(stop\_rewind(\sigma, a, p) \;\wedge\; \big(\sigma^a.next = (-, a+1)\big) \;\wedge\; advance\_pc(\sigma^a, \tau^b) \;\wedge\;$$
$$\tau^b.instr = \langle pfield := q\rangle \;\wedge\;$$
$$default_{avail, active, key, d, isnil, active\_child}(\sigma^a, \tau^{b-1}, \tau^b)\Big).$$

The predicate *step_assgn_nil_to_field*, corresponding to the statements of the form $p \rightarrow pfield := \mathbf{nil}$, can be obtained from *step_assgn_to_field* by replacing the symbolic instruction $\langle pfield := q\rangle$ with $\langle pfield := \mathbf{nil}\rangle$.

Next, we deal with the two instructions that write or read the data field of a node.

$$\underline{step\_assgn\_to\_data}(\sigma, a, \tau, b, p, exp) \stackrel{\text{def}}{=} find\_or\_fail(\sigma, a, \tau, b, p) \ \vee \qquad\qquad p{\to}\text{data} := exp$$

$p$ points to the current node:

$$\Big( stop\_rewind(\sigma, a, p) \ \wedge \ \big(\sigma^a.next = (-, a+1)\big) \ \wedge \ advance\_pc(\sigma^a, \tau^b) \ \wedge$$

$$\tau^b.key = exp\big[d \mapsto \sigma^a.d\big]_{d \in DV_P} \ \wedge$$

$$default_{avail,active,d,isnil,instr,active\_child}(\sigma^a, \tau^{b-1}, \tau^b) \Big).$$

$$\underline{step\_assgn\_to\_var}(\sigma, a, \tau, b, d, p) \stackrel{\text{def}}{=} find\_or\_fail(\sigma, a, \tau, b, p) \ \vee \qquad\qquad d := p{\to}\text{data}$$

$p$ points to the current node:

$$\Big( stop\_rewind(\sigma, a, p) \ \wedge \ \big(\sigma^a.next = (-, a+1)\big) \ \wedge \ advance\_pc(\sigma^a, \tau^b) \ \wedge$$

$$\tau^b.d = \sigma^a.key \ \wedge \ \bigwedge_{d' \in DV_P \setminus \{d\}}(\tau^b.d' = \sigma^a.d') \ \wedge$$

$$default_{avail,active,key,isnil,instr,active\_child}(\sigma^a, \tau^{b-1}, \tau^b) \Big).$$

The following predicate handles the deallocation of the node pointed by a given pointer.

$$\underline{step\_free}(\sigma, a, \tau, b, p) \stackrel{\text{def}}{=} find\_or\_fail(\sigma, a, \tau, b, p) \ \vee \qquad\qquad free\ p$$

$p$ points to the current node:

$$\Big( stop\_rewind(\sigma, a, p) \ \wedge \ \big(\sigma^a.next = (-, *)\big) \ \wedge \ advance\_pc(\sigma^a, \tau^b) \ \wedge$$

$$\neg\tau^b.active \ \wedge$$

$$\bigwedge_{q \in PV_P} \Big( \big(points\_here(\sigma, a, q) \to \tau^b.isnil_q\big) \ \wedge$$

$$\big(\neg points\_here(\sigma, a, q) \to \tau^b.isnil_q = \sigma^a.isnil_q\big) \Big) \ \wedge$$

$$default_{avail,key,d,instr,active\_child}(\sigma^a, \tau^{b-1}, \tau^b) \Big).$$

The following auxiliary predicates handle the search for the current value of a pointer variable $p$, including issuing an error if $p$ is **nil**.

$$\underline{find\_or\_fail}(\sigma, a, \tau, b, p) \stackrel{\text{def}}{=} \Big(\sigma^a.isnil_p \ \wedge \ error(\sigma^a, \tau^b)\Big) \vee rewind(\sigma, a, \tau, b, p)$$

$$\underline{error}(f_1, f_2) \stackrel{\text{def}}{=} \big(f_1.next = (-, *)\big) \ \wedge \ \big(f_2.instr = \text{ERR}\big) \ \wedge$$

$$default_{avail,active,key,d,isnil,pc,active\_child}(f_1, f_1, f_2).$$

Next, we move to the instruction that assigns to a Boolean variable the result of the comparison between two pointers. If at least one of the two pointers is **nil**, the comparison can be resolved locally. Otherwise, a rewind operation may be necessary. The auxiliary predicates *rewind2*, *stop_rewind2*, and *are_equal_after_rewind* are described in Sec. C.7.

$$\underline{step\_assgn\_cond}(\sigma, a, \tau, b, d_{\text{bool}}, p, q) \overset{\text{def}}{=} \qquad\qquad d_{\text{bool}} := (p = q)$$

Case 1: at least one pointer is nil

$$\Big( \big(\sigma^a.isnil_p \vee \sigma^a.isnil_q\big) \wedge$$
$$\big(\sigma^a.next = (-, *)\big) \wedge advance\_pc(\sigma^a, \tau^b) \wedge$$
$$\tau^b.d_{\text{bool}} = \big(\sigma^a.isnil_p \leftrightarrow \sigma^a.isnil_q\big) \wedge$$
$$\textstyle\bigwedge_{d \in DV_P \setminus \{d_{\text{bool}}\}} (\tau^b.d = \sigma^a.d) \wedge$$
$$default_{avail, active, key, isnil, instr, active\_child}\big(\sigma^a, \tau^{b-1}, \tau^b\big) \Big) \quad \vee$$

Case 2: start or keep rewinding

$$rewind2(\sigma, a, \tau, b, p, q) \quad \vee$$

Case 3: stop rewinding

$$\Big( stop\_rewind2(\sigma, a, p, q) \wedge \big(\sigma^a.next = (-, *)\big) \wedge advance\_pc(\sigma^a, \tau^b) \wedge$$
$$\tau^b.d_{\text{bool}} = are\_equal\_after\_rewind(\sigma, a, p, q)\big) \wedge$$
$$\textstyle\bigwedge_{d \in DV_P \setminus \{d_{\text{bool}}\}} (\tau^b.d = \sigma^a.d) \wedge$$
$$default_{avail, active, key, isnil, instr, active\_child}\big(\sigma^a, \tau^{b-1}, \tau^b\big) \Big).$$

The most complex walking instruction is the assignment of the form $p := q \rightarrow pfield$, because it may involve *two* consecutive rewinding phases, as explained in Section 4.

$$\underline{step\_assgn\_from\_field}(\sigma, a, \tau, b, p, q, pfield) \overset{\text{def}}{=} \qquad\qquad p := q \rightarrow pfield$$

Case 1: $q$ is nil; null pointer dereference

$$\big(\sigma^a.isnil_q \wedge error(\sigma^a, \tau^b)\big) \quad \vee$$

Case 2a: phase I; $q$ points elsewhere; start or keep rewinding

$$\big(\sigma^a.instr \neq \text{RWD}_{*,*} \wedge rewind(\sigma, a, \tau, b, q)\big) \quad \vee$$

Case 2b: phase II; $r$ points elsewhere; start or keep rewinding

$$\big(\sigma^a.instr = \text{RWD}_{i,r} \wedge rewind\_special(\sigma, a, \tau, b, r, i)\big) \quad \vee$$

Case 3a: end of phase I; $q$ points to the current node

$$\Big( \sigma^a.instr \neq \text{RWD}_{*,*} \wedge stop\_rewind(\sigma, a, q) \wedge$$
$$\Big(\big(is\_pfield\_nil(\sigma, a, pfield) \vee$$
$$(is\_pfield\_implicit(\sigma, a, pfield) \wedge \neg\sigma^a.active\_child_{index(pfield)})\big) \rightarrow$$
$$step\_assign\_nil(\sigma^a, \tau^b, p)\Big) \wedge$$
$$\Big(\big(is\_pfield\_implicit(\sigma, a, pfield) \wedge \sigma^a.active\_child_{index(pfield)}\big) \rightarrow$$
$$\big(\sigma^a.next = (index(pfield), b) \wedge advance\_pc(\sigma^a, \tau^b) \wedge$$

$$\tau^b.instr = \langle p \coloneqq \mathbf{here} \rangle \ \wedge$$

$$default_{avail,active,key,d,isnil,active\_child}(\sigma^a, \tau^{b-1}, \tau^b)) \Big) \ \wedge$$

$$\bigwedge_{r \in PV_P, i \in [2,n]} \Big( \big( is\_pfield\_ptr(\sigma, a, pfield, r, i) \wedge points\_here(\sigma, i, r) \big) \rightarrow$$

$$set\_ptr\_here(\sigma^a, \tau^b, p) \Big) \ \wedge$$

$$\bigwedge_{r \in PV_P, i \in [2,n]} \Big( \big( is\_pfield\_ptr(\sigma, a, pfield, r, i) \wedge \neg points\_here(\sigma, i, r) \big) \rightarrow$$

$$rewind\_special(\sigma, a, \tau, b, r, i) \Big) \Big) \quad \vee$$

Case 3b: end of phase II; $r$ points to the current node

$$\Big( \sigma^a.instr = \textsc{rwd}_{i,r} \wedge points\_here(\sigma, i, r) \ \wedge \ set\_ptr\_here(\sigma^a, \tau^b, p) \Big).$$

The following auxiliary predicates check the value of the node field $pfield$ at $(\sigma, a)$.

$pfield$ is $\mathbf{nil}$ at the frame $(\sigma, a)$ of the lace:

$$\underline{is\_pfield\_nil}(\sigma, a, pfield) \overset{\text{def}}{=}$$

$$\bigvee_{i \in [2,a]} \Big( \sigma^i.instr = \langle pfield \coloneqq \mathbf{nil} \rangle \wedge \bigwedge_{j \in [i+1,a]} \sigma^j.instr \neq \langle pfield \coloneqq * \rangle \Big)$$

$pfield$ has the same value as $r$ at the frame $(\sigma, a)$ of the lace:

$$\underline{is\_pfield\_ptr}(\sigma, a, pfield, r, i) \overset{\text{def}}{=}$$

$$\sigma^i.instr = \langle pfield \coloneqq r \rangle \wedge \bigwedge_{j \in [i+1,a]} \sigma^j.instr \neq \langle pfield \coloneqq * \rangle$$

$pfield$ has never been assigned up to the frame $(\sigma, a)$ of the lace:

$$\underline{is\_pfield\_implicit}(\sigma, a, pfield) \overset{\text{def}}{=} \bigwedge_{j \in [2,a]} \sigma^j.instr \neq \langle pfield \coloneqq * \rangle \ .$$

## C.6   Boolean Conditions and Control Flow Instructions

Control-flow statements **if** and **while** have two successors, depending on the value of their Boolean condition. To support those statements, we introduce a 3-argument version of the predicate that advances the program counter:

$$\underline{advance\_pc}(f_1, f_2, cond) \overset{\text{def}}{=} \big( f_2.pc = succ(f_1.pc, cond) \big).$$

The next predicate handles the Boolean conditions of the form $p = q$:

$$\underline{step\_cmp\_ptr}(\sigma, a, \tau, b, p, q, neg) \overset{\text{def}}{=} \qquad\qquad\qquad\qquad\qquad \text{if } p = q$$

Case 1: at least one pointer is **nil**

$$\Big( (\sigma^a.isnil_p \vee \sigma^a.isnil_q) \ \wedge \ (\sigma^a.next = (-, a+1)) \ \wedge$$
$$advance\_pc(\sigma^a, \tau^b, \mathrm{xor}(neg, \sigma^a.isnil_p \leftrightarrow \sigma^a.isnil_q)) \ \wedge$$
$$default_{avail,active,key,d,isnil,instr,active\_child}(\sigma^a, \tau^{b-1}, \tau^b) \Big) \quad \vee$$

Case 2: start or keep rewinding

$$rewind2(\sigma, a, \tau, b, p, q) \quad \vee$$

Case 3: stop rewinding

$$\Big( stop\_rewind2(\sigma, a, p, q) \ \wedge \ (\sigma^a.next = (-, a+1)) \ \wedge$$
$$advance\_pc(\sigma^a, \tau^b, \mathrm{xor}(neg, are\_equal\_after\_rewind(\sigma, a, p, q))) \ \wedge$$
$$default_{avail,active,key,d,isnil,instr,active\_child}(\sigma^a, \tau^{b-1}, \tau^b) \Big).$$

The next predicate handles the Boolean conditions based on the content of the data variables. Those conditions can always be resolved locally.

$$\underline{step\_local\_branch}(f_1, f_2, r, exp_1, \ldots, exp_l) \overset{\text{def}}{=} \qquad\qquad\qquad\qquad \text{if}$$
$$\qquad\qquad\qquad\qquad\qquad\qquad\qquad\qquad\qquad\qquad\qquad\qquad\qquad r(exp_1, \ldots, exp_l)$$
$$(f_1.next = (-, *)) \ \wedge$$
$$advance\_pc\big(f_1, f_2, r(exp_1[d \mapsto f_1.d]_{d \in DV_P}, \ldots, exp_l[d \mapsto f_1.d]_{d \in DV_P})\big) \ \wedge$$
$$default_{avail,active,key,d,isnil,instr,active\_child}(f_1, f_2).$$

## C.7 Rewinding the Lace

The following predicate is true when the pointer $q$ is not **nil** at $(\sigma, a)$ and its value cannot be ascertained by analyzing the label $\sigma$ alone. In that case, it is necessary to perform at least one rewinding step, represented by the frame $(\tau, b)$.

$$\underline{rewind}(\sigma, a, \tau, b, q) \overset{\text{def}}{=}$$
$$\sigma^a.next = (dir, b) \ \wedge \ \neg\sigma^a.isnil_q \ \wedge \ a' = cur\_rewind\_pos(\sigma, a) \ \wedge$$
$$\neg points\_here(\sigma, a', q) \ \wedge \ a'' = last\_upd(\sigma, a', q) \ \wedge$$
$$\sigma^{a''}.prev = (dir, b') \ \wedge \ \tau^{b'}.next = (dir', a'') \ \wedge$$
$$b = last\_visit(\tau, dir', a') + 1 \ \wedge \ \tau^b.instr = \mathrm{RWD}_{b'} \ \wedge$$
$$default_{avail,active,pc,key,d,isnil,active\_child}(\sigma^a, \tau^{b-1}, \tau^b).$$

The above predicate uses several auxiliary functions and predicates. The function $cur\_rewind\_pos(\sigma, a)$ returns the position within $\sigma$ of the current rewinding. If the rewinding is just starting, this position will simply be $a$. Otherwise, the current instruction will be of the type $\mathrm{RWD}_b$, and the current rewinding position will be $b$.

$$\underline{cur\_rewind\_pos}(\sigma, a) \overset{\text{def}}{=} \begin{cases} b & \text{if } \sigma^a.instr = \mathrm{RWD}_b \text{ for some } b \in [n], \\ a & \text{otherwise.} \end{cases}$$

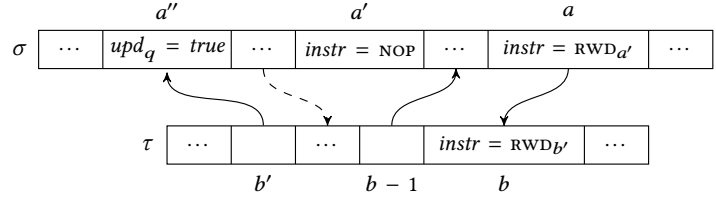

Fig. 9. An example of the $rewind(\sigma, a, \tau, b, q)$ predicate. Arrows connect a frame to its successor in the lace.

The predicate $points\_here(\sigma, a, q)$ holds when $q$ points to the node labeled with $\sigma$ when the lace is at $(\sigma, a)$. This occurs when an instruction of the type $\langle q := \mathbf{here} \rangle$ is found in $\sigma$ at position at most $a$, and all $upd_q$ flags from that position to position $a$ are false.

$$\underline{points\_here}(\sigma, a, q) \overset{\text{def}}{=} \bigvee_{i \in [2,a]} \left( \sigma^i.instr = \langle q := \mathbf{here} \rangle \ \wedge \bigwedge_{j \in [i+1,a]} \neg \sigma^j.upd_q \right).$$

When the value of pointer $q$ cannot be ascertained by the current label $\sigma$, we search for its most recent assignment by identifying the latest position in $\sigma$ that comes before position $a$ and contains the flag $upd_q = true$. This is the job of the function $last\_upd(\sigma, a, q)$. If no position in $\sigma$ before $a$ contains $upd_q = true$, this function returns 2, because in that case the rewinding operation must go back to the frame that lead to the first visit to the current node.

$$\underline{last\_upd}(\sigma, a, q) \overset{\text{def}}{=} \begin{cases} \max\{i \in [2, a] \mid \sigma^i.upd_q = true\} & \text{if } \{i \in [2, a] \mid \sigma^i.upd_q = true\} \neq \emptyset \\ 2 & \text{otherwise.} \end{cases}$$

The last auxiliary function, $last\_visit(\tau, dir, j)$, returns the position of the last frame from $\tau$ whose $next$ field is of the form $(dir, c)$ for some $c \leq j$.

$$\underline{last\_visit}(\tau, dir, j) \overset{\text{def}}{=} \max\{i \in [n] \mid \tau^i.next = (dir, c), c \leq j\}.$$

### C.7.1 Searching for Two Pointers.
We describe a variant of the predicate $rewind$, for the instructions that search for *two* different pointers at the same time. In fact, this only happens in the encoding of the Boolean condition $q_1 = q_2$.

$$\begin{aligned}
\underline{rewind2}&(\sigma, a, \tau, b, q_1, q_2) \overset{\text{def}}{=} \\
& \sigma^a.next = (dir, b) \ \wedge \ \neg\sigma^a.isnil_{q_1} \ \wedge \ \neg\sigma^a.isnil_{q_2} \ \wedge \\
& a' = cur\_rewind\_pos(\sigma, a) \ \wedge \\
& \neg points\_here(\sigma, a', q_1) \ \wedge \ \neg points\_here(\sigma, a', q_2) \ \wedge \\
& a'' = \max\left\{ last\_upd(\sigma, a', q_1), last\_upd(\sigma, a', q_2) \right\} \ \wedge \\
& \sigma^{a''}.prev = (dir, b') \ \wedge \ \tau^{b'}.next = (dir', a'') \ \wedge \\
& b = last\_visit(\tau, dir', a') + 1 \ \wedge \ \tau^b.instr = \text{RWD}_{b'} \ \wedge \\
& default_{avail,active,pc,key,d,isnil,active\_child}(\sigma^a, \tau^{b-1}, \tau^b).
\end{aligned}$$

Next, the version of rewinding used by the second phase of the statement $p := q \rightarrow pfield$.

$$\underline{rewind\_special}(\sigma, a, \tau, b, r, a') \stackrel{\text{def}}{=}$$

$$\sigma^a.next = (dir, b) \; \wedge \; \neg\sigma^a.isnil_r \; \wedge$$

$$\neg points\_here(\sigma, a', r) \; \wedge \; a'' = last\_upd(\sigma, a', r) \; \wedge$$

$$\sigma^{a''}.prev = (dir, b') \; \wedge \; \tau^{b'}.next = (dir', a'') \; \wedge$$

$$b = last\_visit(\tau, dir', a') + 1 \; \wedge \; \tau^b.instr = \text{RWD}_{b',r} \; \wedge$$

$$default_{avail,active,pc,key,d,isnil,active\_child}(\sigma^a, \tau^{b-1}, \tau^b) \, .$$

The following predicates check whether the search for one or two pointers is finished because those variables point to the current node (i.e., the node with label $\sigma$):

$$\underline{stop\_rewind}(\sigma, a, q) \stackrel{\text{def}}{=} \neg\sigma^a.isnil_q \; \wedge \; a' = cur\_rewind\_pos(\sigma, a) \; \wedge$$

$$points\_here(\sigma, a', q)$$

$$\underline{stop\_rewind2}(\sigma, a, q_1, q_2) \stackrel{\text{def}}{=} stop\_rewind(\sigma, a, q_1) \vee stop\_rewind(\sigma, a, q_2) \, .$$

The following predicate is used by the instructions that have already searched for two pointers and want to check whether they are equal.

$$\underline{are\_equal\_after\_rewind}(\sigma, a, p, q) \stackrel{\text{def}}{=}$$

$$a' = cur\_rewind\_pos(\sigma, a) \; \wedge \; points\_here(\sigma, a', p) \wedge points\_here(\sigma, a', q) \, .$$